\documentclass[12pt,a4paper]{article}
\usepackage{hyperref}
\hypersetup{%
      colorlinks=false, linktocpage=false, pdfborder={0 0 0}, pdfstartpage=1, pdfstartview=FitV,%
}

\usepackage{ifthen} 
\newboolean{pdflatex}
\setboolean{pdflatex}{true} 

\newboolean{articletitles}
\setboolean{articletitles}{true} 

\newboolean{uprightparticles}
\setboolean{uprightparticles}{false} 

\newboolean{inbibliography}
\setboolean{inbibliography}{false} 

\usepackage{microtype}
\usepackage{lineno}  
\usepackage{xspace} 
\usepackage{caption} 

\usepackage{graphicx}  
\usepackage{color}
\usepackage{colortbl}
\graphicspath{{./figs/}} 

\usepackage{amsmath} 
\usepackage{amssymb}
\usepackage{amsfonts}
\usepackage{upgreek} 

\newcommand*\patchAmsMathEnvironmentForLineno[1]{%
\expandafter\let\csname old#1\expandafter\endcsname\csname #1\endcsname
\expandafter\let\csname oldend#1\expandafter\endcsname\csname
end#1\endcsname
 \renewenvironment{#1}%
   {\linenomath\csname old#1\endcsname}%
   {\csname oldend#1\endcsname\endlinenomath}%
}
\newcommand*\patchBothAmsMathEnvironmentsForLineno[1]{%
  \patchAmsMathEnvironmentForLineno{#1}%
  \patchAmsMathEnvironmentForLineno{#1*}%
}
\AtBeginDocument{%
\patchBothAmsMathEnvironmentsForLineno{equation}%
\patchBothAmsMathEnvironmentsForLineno{align}%
\patchBothAmsMathEnvironmentsForLineno{flalign}%
\patchBothAmsMathEnvironmentsForLineno{alignat}%
\patchBothAmsMathEnvironmentsForLineno{gather}%
\patchBothAmsMathEnvironmentsForLineno{multline}%
\patchBothAmsMathEnvironmentsForLineno{eqnarray}%
}

\usepackage{hyperref}    
\usepackage[all]{hypcap} 

\usepackage{xspace} 
\usepackage{upgreek}

\def\lhcb {\mbox{LHCb}\xspace}

\def\lhc    {\mbox{LHC}\xspace}

\def\MagUp {\mbox{\em Mag\kern -0.05em Up}\xspace}

\ifthenelse{\boolean{uprightparticles}}%
{

 \def\Ppi         {\ensuremath{\uppi}\xspace}

 \def\Ppsi        {\ensuremath{\uppsi}\xspace}

 \def\PDelta      {\ensuremath{\Delta}\xspace}                 
 \def\PXi      {\ensuremath{\Xi}\xspace}                 
 \def\PLambda      {\ensuremath{\Lambda}\xspace}                 
 \def\PSigma      {\ensuremath{\Sigma}\xspace}                 
 \def\POmega      {\ensuremath{\Omega}\xspace}                 
 \def\PUpsilon      {\ensuremath{\Upsilon}\xspace}                 
 
 \def\PB      {\ensuremath{\mathrm{B}}\xspace}                 
                  
 \def\PD      {\ensuremath{\mathrm{D}}\xspace}

 \def\PJ      {\ensuremath{\mathrm{J}}\xspace}                 
 \def\PK      {\ensuremath{\mathrm{K}}\xspace}

 \def\Pb      {\ensuremath{\mathrm{b}}\xspace}

 \def\Pi      {\ensuremath{\mathrm{i}}\xspace}

}
{

 \def\Ppi         {\ensuremath{\pi}\xspace}

 \def\Ppsi        {\ensuremath{\psi}\xspace}                 
                  
 \mathchardef\PDelta="7101
 \mathchardef\PXi="7104
 \mathchardef\PLambda="7103
 \mathchardef\PSigma="7106
 \mathchardef\POmega="710A
 \mathchardef\PUpsilon="7107
                  
 \def\PB      {\ensuremath{B}\xspace}                 
                  
 \def\PD      {\ensuremath{D}\xspace}

 \def\PJ      {\ensuremath{J}\xspace}                 
 \def\PK      {\ensuremath{K}\xspace}

 \def\Pb      {\ensuremath{b}\xspace}

 \def\Pi      {\ensuremath{i}\xspace}

}

\makeatletter
\ifcase \@ptsize \relax
  \newcommand{\miniscule}{\@setfontsize\miniscule{4}{5}}
\or
  \newcommand{\miniscule}{\@setfontsize\miniscule{5}{6}}
\or
  \newcommand{\miniscule}{\@setfontsize\miniscule{5}{6}}
\fi
\makeatother

\DeclareRobustCommand{\optbar}[1]{\shortstack{{\miniscule (\rule[.5ex]{1.25em}{.18mm})}
  \\ [-.7ex] $#1$}}

\def\bquark    {{\ensuremath{\Pb}}\xspace}

\def\pion   {{\ensuremath{\Ppi}}\xspace}

\def\pim    {{\ensuremath{\pion^-}}\xspace}

\def\kaon    {{\ensuremath{\PK}}\xspace}
  \def\Kbar    {{\kern 0.2em\overline{\kern -0.2em \PK}{}}\xspace}

\def\KorKbar    {\kern 0.18em\optbar{\kern -0.18em K}{}\xspace}

\def\Kp      {{\ensuremath{\kaon^+}}\xspace}

  \def\Dbar    {{\kern 0.2em\overline{\kern -0.2em \PD}{}}\xspace}
\def\D       {{\ensuremath{\PD}}\xspace}

\def\DorDbar    {\kern 0.18em\optbar{\kern -0.18em D}{}\xspace}
\def\Dz      {{\ensuremath{\D^0}}\xspace}

\def\Dstar   {{\ensuremath{\D^*}}\xspace}

\def\Bbar    {{\ensuremath{\kern 0.18em\overline{\kern -0.18em \PB}{}}}\xspace}

\def\BorBbar    {\kern 0.18em\optbar{\kern -0.18em B}{}\xspace}

\def\jpsi     {{\ensuremath{{\PJ\mskip -3mu/\mskip -2mu\Ppsi\mskip 2mu}}}\xspace}

  \def\Y#1S{\ensuremath{\PUpsilon{(#1S)}}\xspace}

\def\Lz          {{\ensuremath{\PLambda}}\xspace}
\def\Lbar        {{\ensuremath{\kern 0.1em\overline{\kern -0.1em\PLambda}}}\xspace}
\def\LorLbar    {\kern 0.18em\optbar{\kern -0.18em \PLambda}{}\xspace}

\def\Lb      {{\ensuremath{\Lz^0_\bquark}}\xspace}

\def\to                 {\ensuremath{\rightarrow}\xspace}

\def\CP                {{\ensuremath{C\!P}}\xspace}

\def\AT#1     {\ensuremath{A_{\mathrm{T}}^{#1}}\xspace}           

\def\C#1      {\ensuremath{\mathcal{C}_{#1}}\xspace}                       
\def\Cp#1     {\ensuremath{\mathcal{C}_{#1}^{'}}\xspace}                    
\def\Ceff#1   {\ensuremath{\mathcal{C}_{#1}^{\mathrm{(eff)}}}\xspace}        
\def\Cpeff#1  {\ensuremath{\mathcal{C}_{#1}^{'\mathrm{(eff)}}}\xspace}       
\def\Ope#1    {\ensuremath{\mathcal{O}_{#1}}\xspace}                       
\def\Opep#1   {\ensuremath{\mathcal{O}_{#1}^{'}}\xspace}                    

\newcommand{\tev}{\ifthenelse{\boolean{inbibliography}}{\ensuremath{~T\kern -0.05em eV}\xspace}{\ensuremath{\mathrm{\,Te\kern -0.1em V}}}\xspace}
\newcommand{\gev}{\ensuremath{\mathrm{\,Ge\kern -0.1em V}}\xspace}
\newcommand{\mev}{\ensuremath{\mathrm{\,Me\kern -0.1em V}}\xspace}
\newcommand{\kev}{\ensuremath{\mathrm{\,ke\kern -0.1em V}}\xspace}
\newcommand{\ev}{\ensuremath{\mathrm{\,e\kern -0.1em V}}\xspace}
\newcommand{\gevc}{\ensuremath{{\mathrm{\,Ge\kern -0.1em V\!/}c}}\xspace}
\newcommand{\mevc}{\ensuremath{{\mathrm{\,Me\kern -0.1em V\!/}c}}\xspace}
\newcommand{\gevcc}{\ensuremath{{\mathrm{\,Ge\kern -0.1em V\!/}c^2}}\xspace}
\newcommand{\gevgevcccc}{\ensuremath{{\mathrm{\,Ge\kern -0.1em V^2\!/}c^4}}\xspace}
\newcommand{\mevcc}{\ensuremath{{\mathrm{\,Me\kern -0.1em V\!/}c^2}}\xspace}

\def\m    {\ensuremath{\mathrm{ \,m}}\xspace}

\def\cm   {\ensuremath{\mathrm{ \,cm}}\xspace}

\def\mm   {\ensuremath{\mathrm{ \,mm}}\xspace}

\def\mum  {\ensuremath{{\,\upmu\mathrm{m}}}\xspace}

\def\invfb   {\ensuremath{\mbox{\,fb}^{-1}}\xspace}

\def\ns   {\ensuremath{{\mathrm{ \,ns}}}\xspace}

\def\gsim{{~\raise.15em\hbox{$>$}\kern-.85em
          \lower.35em\hbox{$\sim$}~}\xspace}
\def\lsim{{~\raise.15em\hbox{$<$}\kern-.85em
          \lower.35em\hbox{$\sim$}~}\xspace}

\def\tell1  {TELL1\xspace}
\def\ukl1   {UKL1\xspace}

\def\cfourften     {\ensuremath{\mathrm{ C_4 F_{10}}}\xspace}
\def\cffour        {\ensuremath{\mathrm{ CF_4}}\xspace}

\usepackage{mciteplus}

\usepackage{longtable} 

\def\sumetprev{\ensuremath{\Sigma E_T^{\rm{Prev}}}\xspace}
\def\PbPb{\ensuremath{{\rm PbPb}}\xspace}
\def\pPb{\ensuremath{p{\rm Pb}}\xspace}
\def\Pbp{\ensuremath{{\rm Pb}p}\xspace}

\usepackage{overpic}
\usepackage{wrapfig}

\begin{document}

\renewcommand{\thefootnote}{\fnsymbol{footnote}}
\setcounter{footnote}{1}



\begin{titlepage}
\pagenumbering{roman}

\vspace*{-1.5cm}
\centerline{\large EUROPEAN ORGANIZATION FOR NUCLEAR RESEARCH (CERN)}
\vspace*{0.4cm}
\noindent
\begin{tabular*}{\linewidth}{lc@{\extracolsep{\fill}}r@{\extracolsep{0pt}}}
\ifthenelse{\boolean{pdflatex}}
{\vspace*{-2.7cm}\mbox{\!\!\!\includegraphics[width=.14\textwidth]{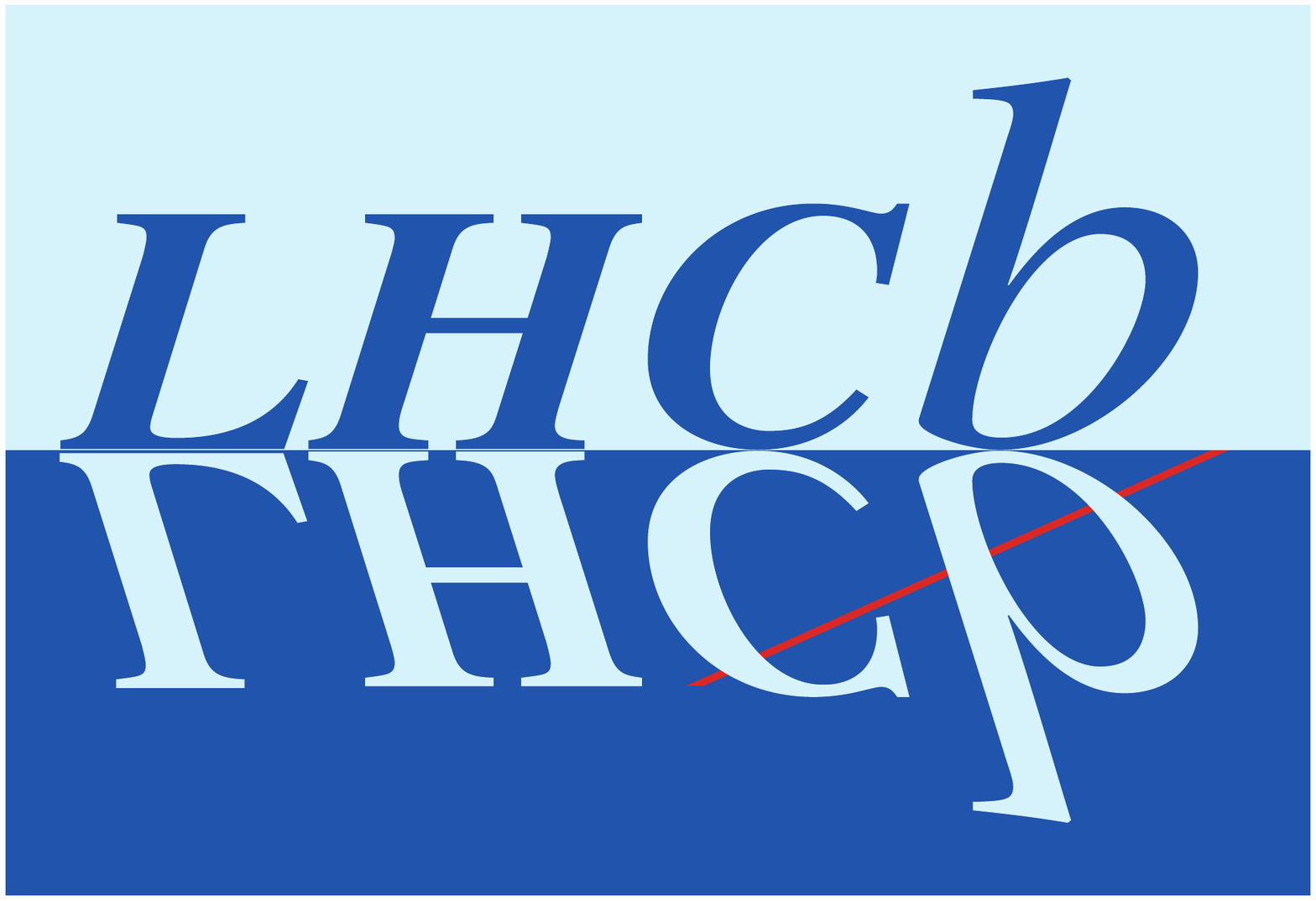}} & &}%
{\vspace*{-1.2cm}\mbox{\!\!\!\includegraphics[width=.12\textwidth]{lhcb-logo.eps}} & &}%
\\
 & & CERN-LHCb-DP-2017-001 \\  
 & & \today \\ 
\end{tabular*}

\vspace*{0.2cm}

{\normalfont\bfseries\boldmath\huge
\begin{center}
  Improved performance\\ of the LHCb Outer Tracker \\in LHC Run 2
  \vspace*{0.1cm}
\end{center}
}

\begin{center}
The LHCb Outer Tracker group
\end{center}


\begin{flushleft}
\center
\small
Ph.~d'Argent$^{	  1}$,
L.~Dufour$^{	  2}$,
L.~Grillo$^{	  7}$,
J.A.~de~Vries$^{  2}$,
A.~Ukleja$^{	  6}$,
R.~Aaij$^{    	  8}$,
F.~Archilli$^{    2}$,
S.~Bachmann$^{	  1}$,
D.~Berninghoff$^{ 1}$,
A.~Birnkraut$^{	  3}$,
J.~Blouw$^{	  9}$\footnotemark[2],
M.~De~Cian$^{	  1}$,
G.~Ciezarek$^{	  2}$,
Ch.~F\"arber$^{	  1}$,
M.~Demmer$^{	  3}$,
F.~Dettori$^{	  4}$,
E.~Gersabeck$^{	  1}$,
J.~Grabowski$^{	  1}$,
W.D.~Hulsbergen$^{2}$,
B.~Khanji$^{	  1}$,
M.~Kolpin$^{	  1}$,
M.~Kucharczyk$^{  5}$,
B.P.~Malecki$^{	  5}$,
M.~Merk$^{	  2}$,
M.~Mulder$^{	  2}$,
J.~M\"uller$^{	  3}$,
V.~Mueller$^{	  3}$,
A.~Pellegrino$^{  2}$,
M.~Pikies$^{	  5}$,
B.~Rachwal$^{	 11}$,
T.~Schmelzer$^{	  3}$,
B.~Spaan$^{	  3}$,
M.~Szczekowski$^{ 6}$,
J.~van~Tilburg$^{ 2}$,
S.~Tolk$^{	 10}$,
N.~Tuning$^{	  2}$,
U.~Uwer$^{	  1}$,
J.~Wishahi$^{     3}$,
M.~Witek$^{	  5}$,

\vspace{0.1cm}

\footnotesize{
$ ^{1}$Physikalisches Institut, Heidelberg, Germany\\
$ ^{2}$Nikhef, Amsterdam, The Netherlands\\
$ ^{3}$Technische Universit\"at Dortmund, Germany\\
$ ^{4}$Oliver Lodge Laboratory, University of Liverpool, Liverpool, United Kingdom\\
$ ^{5}$H. Niewodniczanski Institute of Nuclear Physics, Cracow, Poland\\
$ ^{6}$National Center for Nuclear Research (NCBJ), Warsaw, Poland\\
$ ^{7}$Bicocca University, Milano, Italy\\
$ ^{8}$European Organization for Nuclear Research (CERN), Geneva, Switzerland\\
$ ^{9}$Max Planck-Institut F\"ur Kernphysik (MPIK), Heidelberg, Germany\\
$ ^{10}$Cavendish Laboratory, University of Cambridge, Cambridge, United Kingdom\\
$ ^{11}$AGH - University of Science and Technology, Faculty of Physics and Applied Computer Science, Krakow, Poland\\
\footnotemark[2] Deceased 
}
\vspace{0.1cm}

\end{flushleft}


\begin{abstract}
  \noindent
The LHCb Outer Tracker is a gaseous detector covering an area
of $5\times6\m^2$ with 12 double layers of straw tubes. 
The performance of the detector is presented 
based on data of the LHC Run 2 running period from 2015 and 2016.
Occupancies and operational experience for data collected in $p p$, 
\pPb and \PbPb collisions are described. 
An updated study of the ageing effects is presented 
showing no signs of gain deterioration or other radiation damage effects.
In addition several improvements with respect to LHC Run 1 data taking 
are introduced. A novel real-time calibration of the time-alignment 
of the detector and the alignment of the single monolayers composing detector modules
are presented, improving the drift-time and position resolution of the detector by 20\%.
Finally, a potential use of the improved resolution for the timing of charged tracks
is described, showing the possibility to identify low-momentum hadrons with their time-of-flight.
\end{abstract}


\begin{center}
Published in JINST 12 (2017) no.11, P11016
\end{center}


{\footnotesize 
\centerline{\copyright~CERN on behalf of the \lhcb collaboration, licence \href{http://creativecommons.org/licenses/by/4.0/}{CC-BY-4.0}.}}

\end{titlepage}


\newpage
\setcounter{page}{2}
\mbox{~}
%

\cleardoublepage


\renewcommand{\thefootnote}{\arabic{footnote}}
\setcounter{footnote}{0}

\tableofcontents
\cleardoublepage
\null\cleardoublepage


\pagestyle{plain} 
\setcounter{page}{1}
\pagenumbering{arabic}



\section{Introduction}
\label{sec:Introduction}

The LHCb experiment is dedicated to the study of \CP violation and rare decays of
hadrons with $b$ and $c$ quarks at the Large Hadron Collider.  The LHCb
detector~\cite{Alves:2008zz} is a single arm forward spectrometer, with an
acceptance in the pseudorapidity range $2<\eta<5$.  The tracking system of the
detector is composed of a silicon-strip vertex detector, close to the
proton-proton interaction region, and five tracking stations: two upstream and
three downstream of a dipole magnet with bending power of around 4 Tm.  The
upstream tracking stations and the inner part of the downstream ones are composed
of silicon-strip detectors.  The outer part of the downstream stations is
covered by the Outer Tracker~(OT)~\cite{TDR}, a straw-tube gaseous detector.
The single hit resolution in the OT leads to a momentum resolution of $\delta p/p \approx 0.4\%$,
providing a superior mass resolution of reconstructed charm and bottom decays.

\begin{figure}[!t]
\begin{center}
    \begin{picture}(250,300)(0,0)
    \put(-105,205){\includegraphics[bb=0 0 1982 875,scale=0.12]{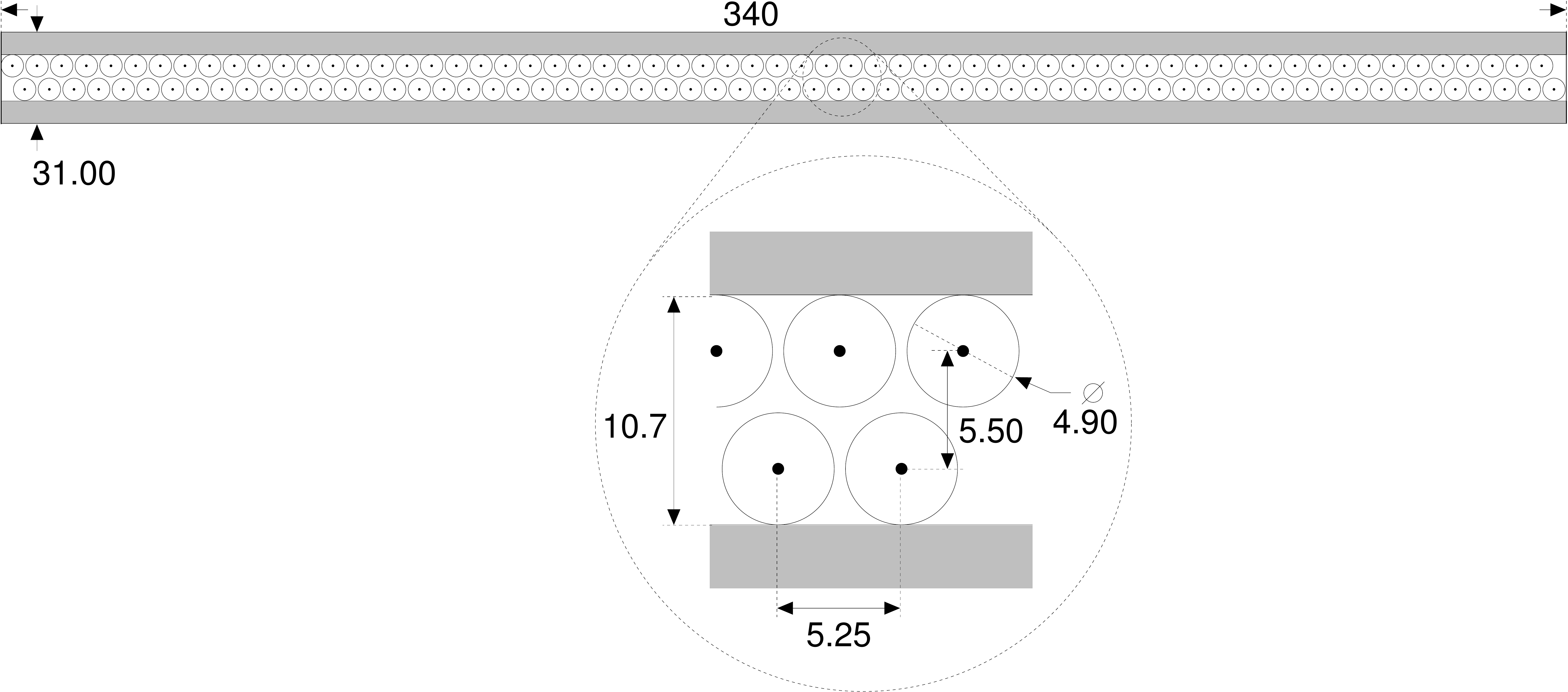}}
    \put(70, 100){\includegraphics[bb=60 127 469 469,scale=0.75]{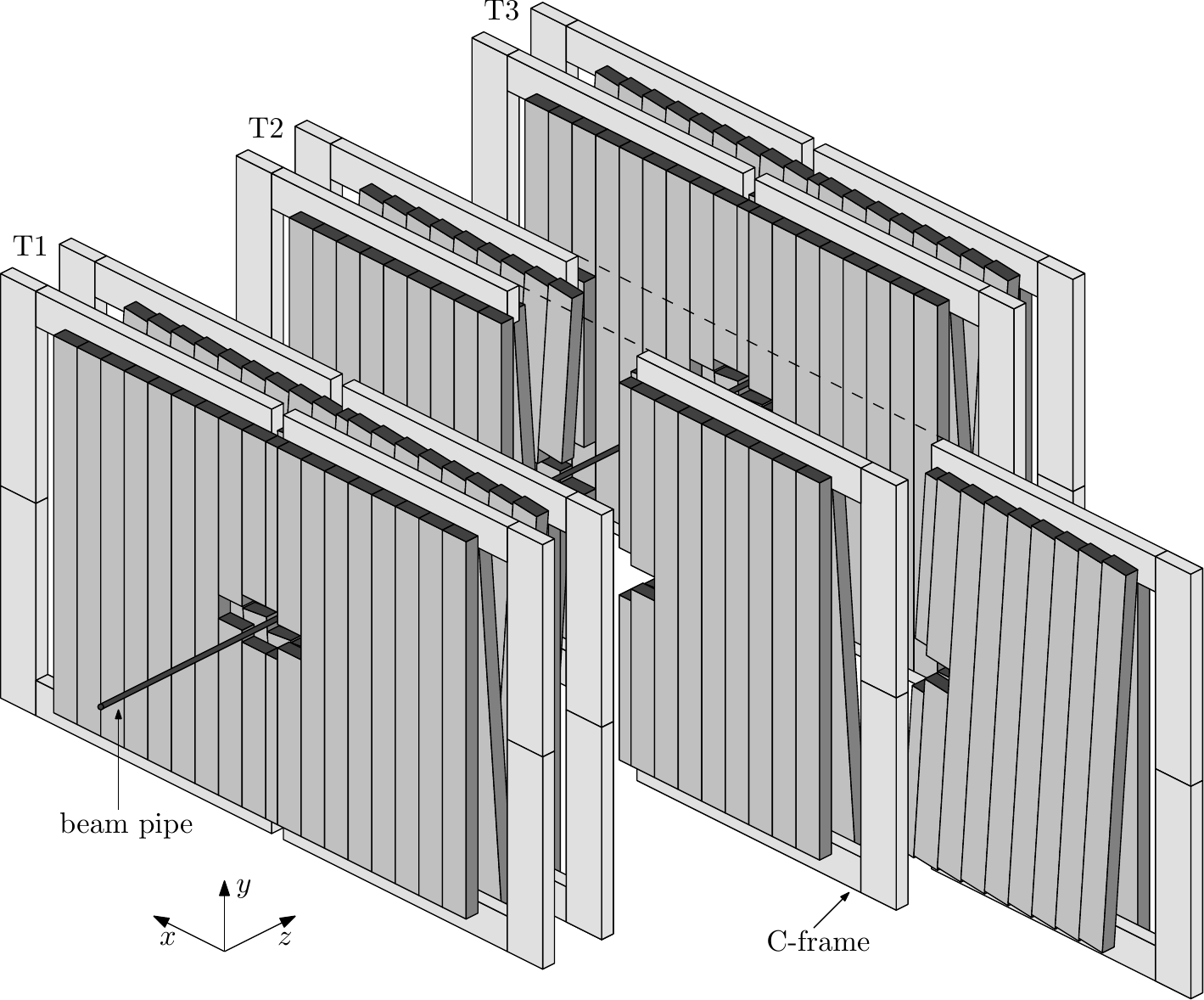}}
    \put(-100,200){(a)}
    \put(30,10){(b)}
    \end{picture}
    \caption{\small 
    (a) OT module cross section, with distances in mm.
    (b) Arrangement of OT straw-tube modules in layers and stations.}
    \label{fig:mod}
\end{center}
\end{figure}

The OT covers an area of about  $5\times6\m^2$
starting from a minimum distance of about 10 cm from the beam direction.
The cross-shaped inner region not covered by the OT, 
$|y|<10\cm$ for $|x|<59.7 \cm$, and $|y|< 20\cm$ for $|x|< 25.6 \cm$,  
is instrumented with silicon strip detectors~\cite{Alves:2008zz}.\footnote{The LHCb coordinate system is a 
right-handed coordinate system, with the z axis pointing along the
beam axis, y the vertical direction, and x the horizontal direction. 
The (x, z) plane is the bending plane
of the dipole magnet.}
Each straw tube measures the drift time of charges created by the ionising particles, 
and collected on the anode wire.  
The straw tubes are $2.4 \m$ long with $4.9 \mm$ inner diameter. 
The anode wire at the straw centre is set to +1550\,V and is made of 
$25 \mum$ thick gold plated tungsten.
The cathode consists of a $40 \mum$ thick inner
foil of electrically conducting carbon doped Kapton-XC~\footnote{Kapton\textregistered~ is 
a polyimide film developed by DuPont.}  
and a $25 \mum$ thick outer foil, consisting of Kapton-XC laminated together
with a $12.5 \mum$ thick layer of aluminium.  
The straws are arranged in modules in two staggered layers glued to sandwich
panels, using Araldite AY103-1.\footnote{Araldite\textregistered~ is a two
component epoxy resin developed by Huntsman.}  Two panels are sealed with
$400 \mum$ thick carbon fibre sidewalls, resulting in a gas-tight box enclosing
a stand-alone detector module.  The straws are filled with a gas mixture of 
Ar/CO$_2$/O$_2$ (70/28.5/1.5) 
which guarantees a fast drift time below $50 \ns$. 
The gas in input to the modules has an overpressure of 2.5~mbar. 
The modules are composed of two staggered layers (monolayers) of 64 drift tubes
each. A cross-section of the module layout is shown in
figure~\ref{fig:mod}(a).

 Two types of modules are used: type $F$ modules, which have an active length 
of 4850\,mm, have monolayers split in the middle into two independent readout sections 
read out from the outer ends;  type $S$ modules have about half the length of $F$-type 
modules, they are mounted above and below the beampipe, and
are read out from the outer module end.    One detector layer is built from 14 long
and 8 short modules, see figure~\ref{fig:mod}(b).  The complete OT detector
consists of 168 long and 96 short modules and comprises 53,760 single straw-tube
channels.

The detector modules are arranged in three stations.  Each station consists of
four module layers, arranged in an {\it x-u-v-x} geometry: the modules in the
$x$-layers are oriented vertically, whereas those in the $u$ and $v$ layers are
tilted by $+5^o$ and $-5^o$ with respect to the vertical, respectively.
This leads to a total of 24 straw layers positioned along the $z$-axis. 

Each station is split into two halves, retractable on both sides of the beam
line. Each half consists of two independently movable units, known as  
C-frames, see figure~\ref{fig:mod}(b).  
The modules are positioned on the C-frames by means of precision
dowel pins.  The C-frames also provide routing for all detector services (gas,
low and high voltage, cooling water, data fibres, slow and fast control).  The
C-frames are sustained by a stainless steel structure (called ``bridge''), equipped
with rails allowing the independent movement of all twelve C-frames. 
The bridge is mounted on a concrete structure (called ``table'').

The front-end (FE) electronics measures the arrival times of the ionization
clusters produced by charged particles traversing the straw-tubes with respect
to the beam crossing (BX) signal~\cite{Berkien:2005zz}.  
The signals are amplified and subsequently discriminated against a threshold value.
This value is set as low as possible to minimize the effect of time-walk on the drift time 
resolution, and at the same time to keep the number of noise hits at an acceptable level.
The measured arrival times are then
digitized for each 25\,ns (the LHC design value for the minimum bunch crossing
interval) and stored in a digital pipeline to await the lowest-level trigger
(L0) decision.  
This TDC value is given in units of 0.4~ns, as the 25~ns window is encoded with 6 bits.
On a positive L0 decision, the digitized data 
of the triggered bunch crossing and of the following two bunch crossings (for a total window of 75 ns) 
are transmitted via optical links to TELL1 boards in the LHCb DAQ
system~\cite{Haefeli:2006cv}.

The performance of the OT as studied in detail during Run~1 of LHC (data-taking years 2009-2012)
can be found in  Ref.~\cite{Arink:1629476}, 
together with additional details of the apparatus. 

Thanks to improvements on the LHC magnets during the Long Shutdown 1 (LS1) period in 2013 and 2014, 
the LHC provided collisions at a higher centre-of-mass energy 
of $13\tev$ in Run 2 (versus 7 and $8 \tev$ in Run 1) and proton bunches with 
a time spacing of $25 \ns$ instead of $50 \ns$. 
While these conditions are closer to the design values, they are novel, challenging
circumstances, motivating the documentation of the performance of the detector in this paper. 

This paper describes the detector performance and improvements in the LHC Run~2
period with data collected in 2015 and 2016.
In the first part, the performance with proton-proton ($pp$) and lead-lead (\PbPb) collisions are 
discussed in section~\ref{sec:performance}. In section~\ref{sec:rasnik} a study is presented of the mechanical stability of the OT
obtained through a hardware alignment system (called ``RASNIK'').
In section~\ref{sec:ageing} an updated study of the ageing of the detector is reported. 

Furthermore several improvements to the OT were carried out during LS1 and are presented in the second part.
Owing to an improved online event filter farm, the two levels of the LHCb High Level Trigger (HLT)
were split in two stages, allowing for a full real-time alignment and calibration of the whole detector
used by the second stage of the online event selection~\cite{BORGHI2017560}.
In section~\ref{sec:realtime} a novel real-time calibration of the global time
alignment of the OT with respect to the LHC clock is presented, while in
section~\ref{sec:singlefe} the single front-end alignment and calibration is
discussed.  
An updated calibration of the distance drift-time relation 
is reported in section~\ref{sec:rt-relation}.
The drift-time measurement capabilities of the OT have been used to
measure the time of flight of tracks, which can be exploited for particle
identification and for primary interaction vertex assignment and distinction.
This method was developed and used for the first time in Run 2, and is
presented in section~\ref{sec:tracktimes} together with its performance.

\cleardoublepage
\part{Operational experience}

The monitoring and evaluation of the detector performance is crucial to maintain the
same quality data needed for the physics measurements as in Run 1.  In this first part of
the paper, the performance and stability of the detector operation is presented,
together with studies on possible radiation damage due to the received dose.

\section{Event occupancies in proton and lead collisions}
\label{sec:performance}

\subsection{Performance in Run 2 \texorpdfstring{$pp$}{} collisions}

The performance of the Outer Tracker in Run 1 exceeded expectations in terms of low number of dead and noisy channels,
single hit resolution, and tracking performance, and has been extensively described elsewhere~\cite{Arink:1629476}.

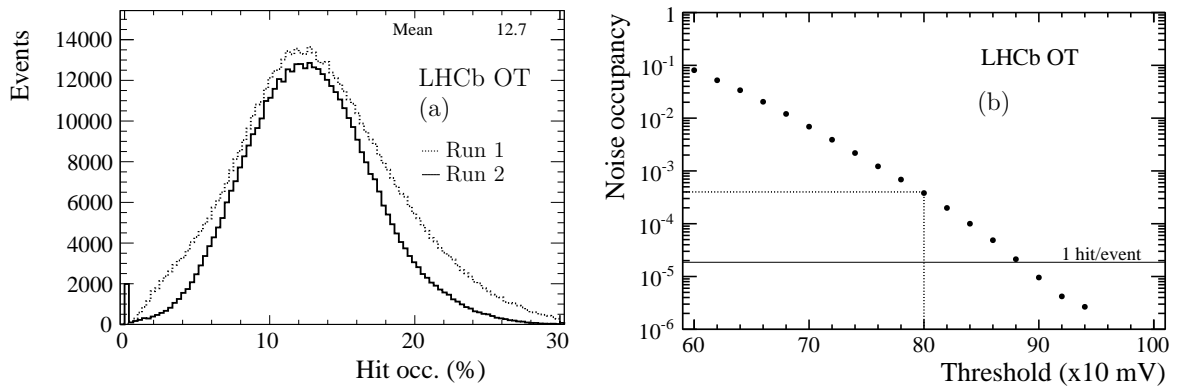
\begin{figure}[!b]
  \begin{center}
   \begin{overpic}[width=0.49\linewidth, trim=1.5cm 1cm 1.5cm 1cm, clip]{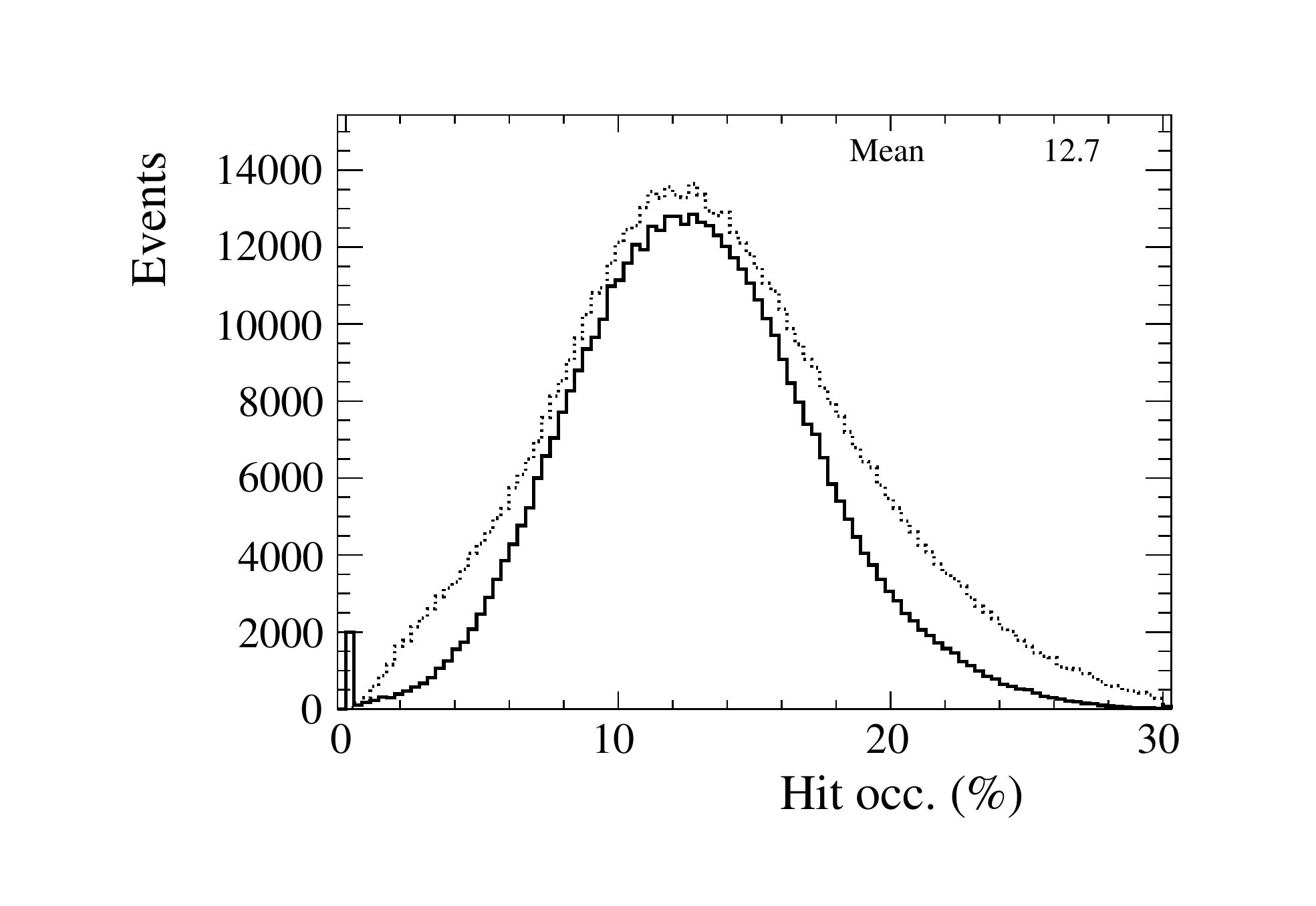}
	\put(70,52){\scalebox{0.8}{ LHCb OT }}
        \put(70,47){\scalebox{0.8}{ (a) }}
        \multiput(72, 41)(0.5, 0){6}{\line(1,0){0.2}}
   	\put(76,40){\scalebox{0.7}{Run 1}}
        \multiput(72, 37)(0.5, 0){6}{\line(1,0){0.5}}
        \put(76,36){\scalebox{0.7}{Run 2}}
    \end{overpic}
   \begin{overpic}[width=0.49\linewidth]{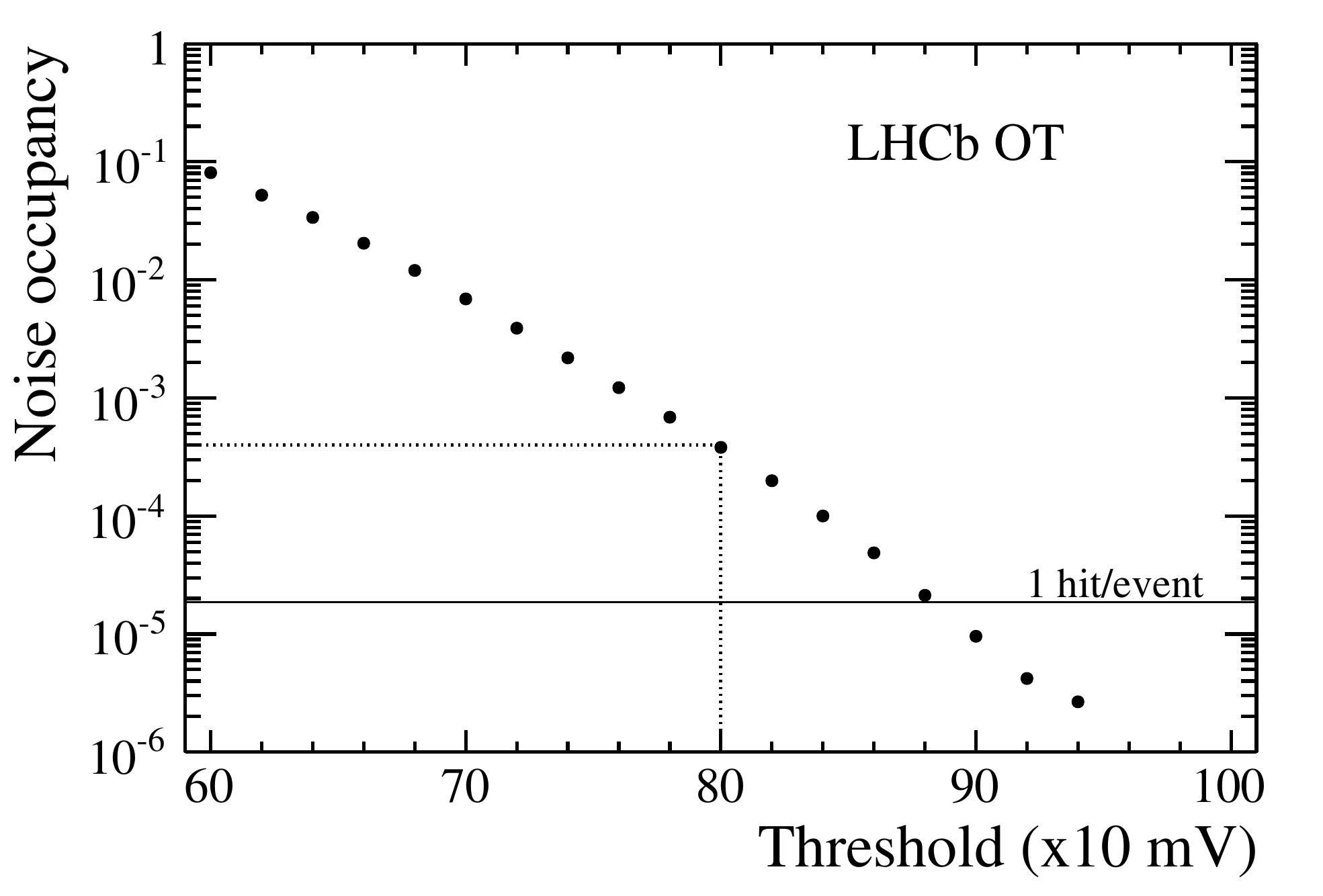}
    \put(62,48){\scalebox{0.8}{ (b) }}
   \end{overpic}
   \end{center}
  \caption{\small (a) Event occupancy for the OT in Run 1 (dottet line) and Run 2 (solid line).  
  The event occupancy in the OT is on average 13\% in Run 2.
(b) Occupancy of noise hits as a function of the discriminator threshold. 
} 
  \label{fig:occ-run12}
\end{figure}

The larger center-of-mass energy and the increased spill-over due to the 25 ns bunch spacing  
lead to a larger OT occupancy.
In order to keep the OT event multiplicity at a similar level with respect to Run 1,
to maintain the same performance of the track reconstruction, the 
average number of overlapping events per bunch crossing was reduced from $\mu = 1.7$ in Run 1,
to $\mu=1.1$ in Run 2. The resulting distributions of the OT event occupancies are compared in
figure~\ref{fig:occ-run12}(a). Due to the fact that the event occupancy in Run 2 is the result 
of adding the occupancies of multiple bunch crossings, the final event occupancy has a smaller spread.
In 2017 the OT hit occupancy was further reduced by limiting the number of spill-over hits from the previous 
bunch crossing (see section~\ref{sec:spillover}).

The fraction of noise hits that contribute to the event occupancy is negligible.
With the nominal discriminator threshold of 800~mV, about 30 noise hits are expected every event, 
as shown in figure~\ref{fig:occ-run12}(b).
As far as dead channels are concerned, only 8 out of a total of 53760 straws 
do not give any signal, as they had to be disconnected to avoid large currents due to short-circuits.

\subsection{Occupancies and hits from previous bunch crossings}
\label{sec:spillover}

The OT readout window contains three consecutive bunch crossings of 25~ns upon each L0 trigger,
to ensure that late hits with long drift-times are also recorded.
Similarly, late hits from the previous bunch crossing will be recorded in the OT as
early hits from the triggered bunch crossing. These so-called spill-over hits
hamper the track reconstruction, enlarging the fraction of fake tracks.
In addition, these hits from the previous spill consume bandwidth in the data transmission.

\begin{figure}[!b]
  \begin{center}
      \begin{overpic}[width=0.49\linewidth]{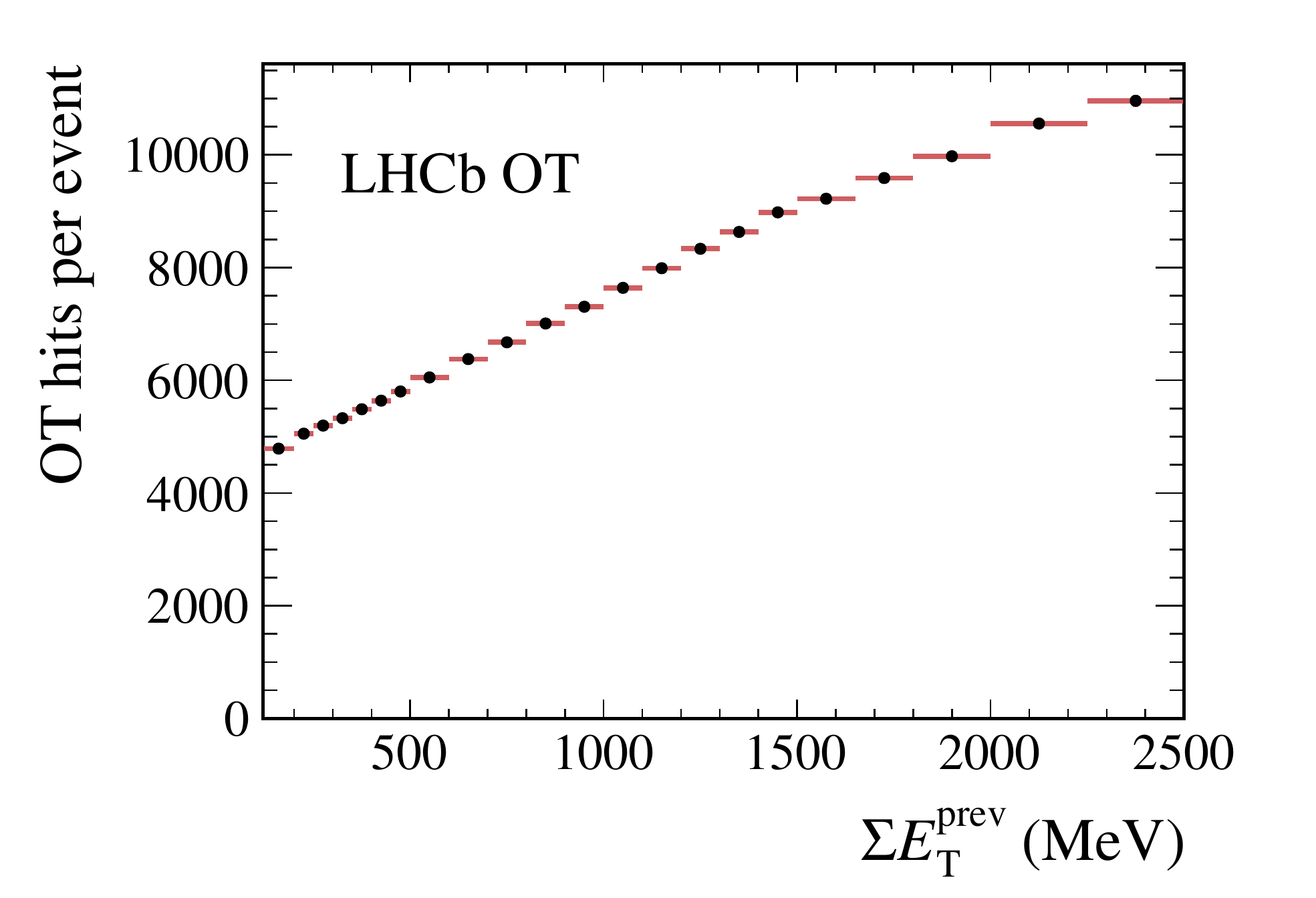}
       \put(26,45){(a)}
      \end{overpic}
    \begin{overpic}[width=0.49\linewidth]{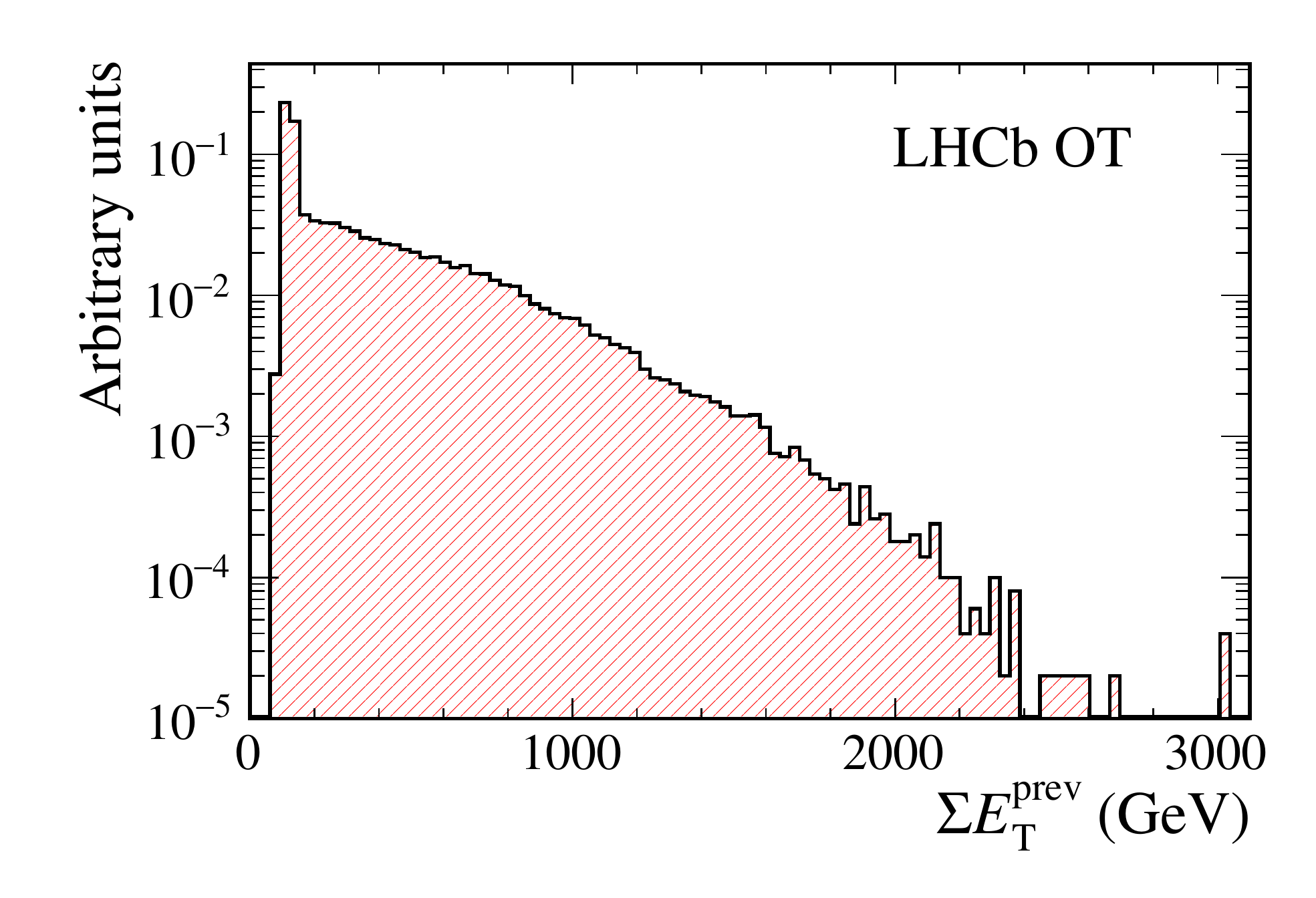}
     \put(68,47	){(b)}
    \end{overpic}
  \end{center}
  \caption{
    \small (a) The average number of recorded hits as a function of the activity in the
    previous bunch crossing, expressed as the scalar sum of transverse energy in all calorimeter clusters, $\Sigma E_T$.
  (b) The distribution of $\Sigma E_T$ for unbiased events, where one unit corresponds to 24~MeV.} 
  \label{fig:performance-sumET}
\end{figure}

The scalar sum of the transverse energy of all hadronic calorimeter clusters of 
the previous bunch crossing,  \sumetprev,  is available in the L0 trigger decision unit. 
To quantify the influence of spill-over on the tracker's performance, the occupancy
and drift times of no-bias events are analysed in bins of the scalar sum
of transverse energy of all hadronic calorimeter clusters in the previous bunch crossing. 
The average number of hits recorded as a function of  \sumetprev
is shown in figure~\ref{fig:performance-sumET}(a). It can be seen that the occupancy in the OT
is double for events with a very busy previous event,  $\sumetprev\approx 2000 (= 48 \gev)$.

\begin{figure}[!t]
  \begin{center}
    \includegraphics[width=0.49\linewidth]{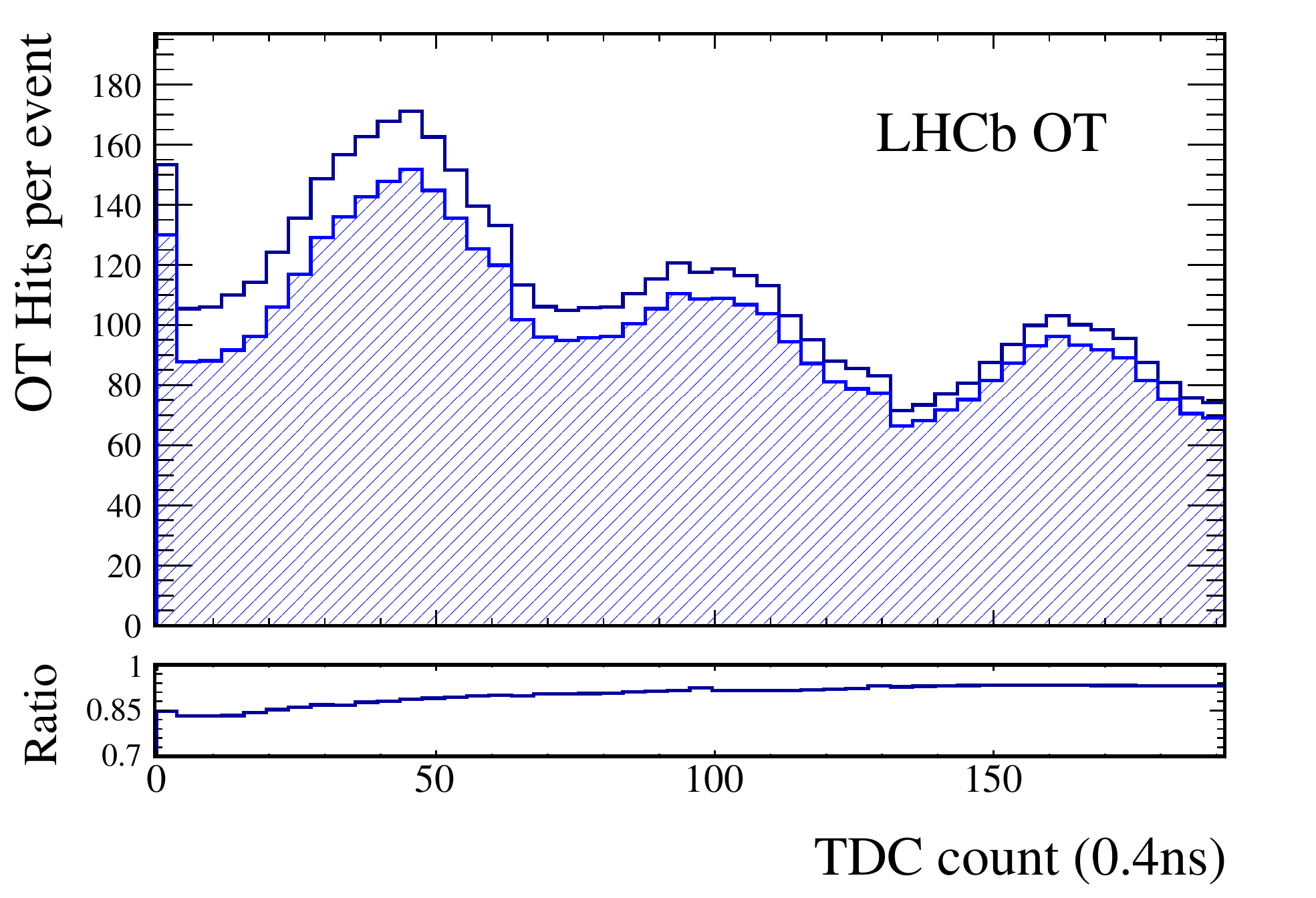}\put(-75,110){(a)}
    \includegraphics[width=0.49\linewidth]{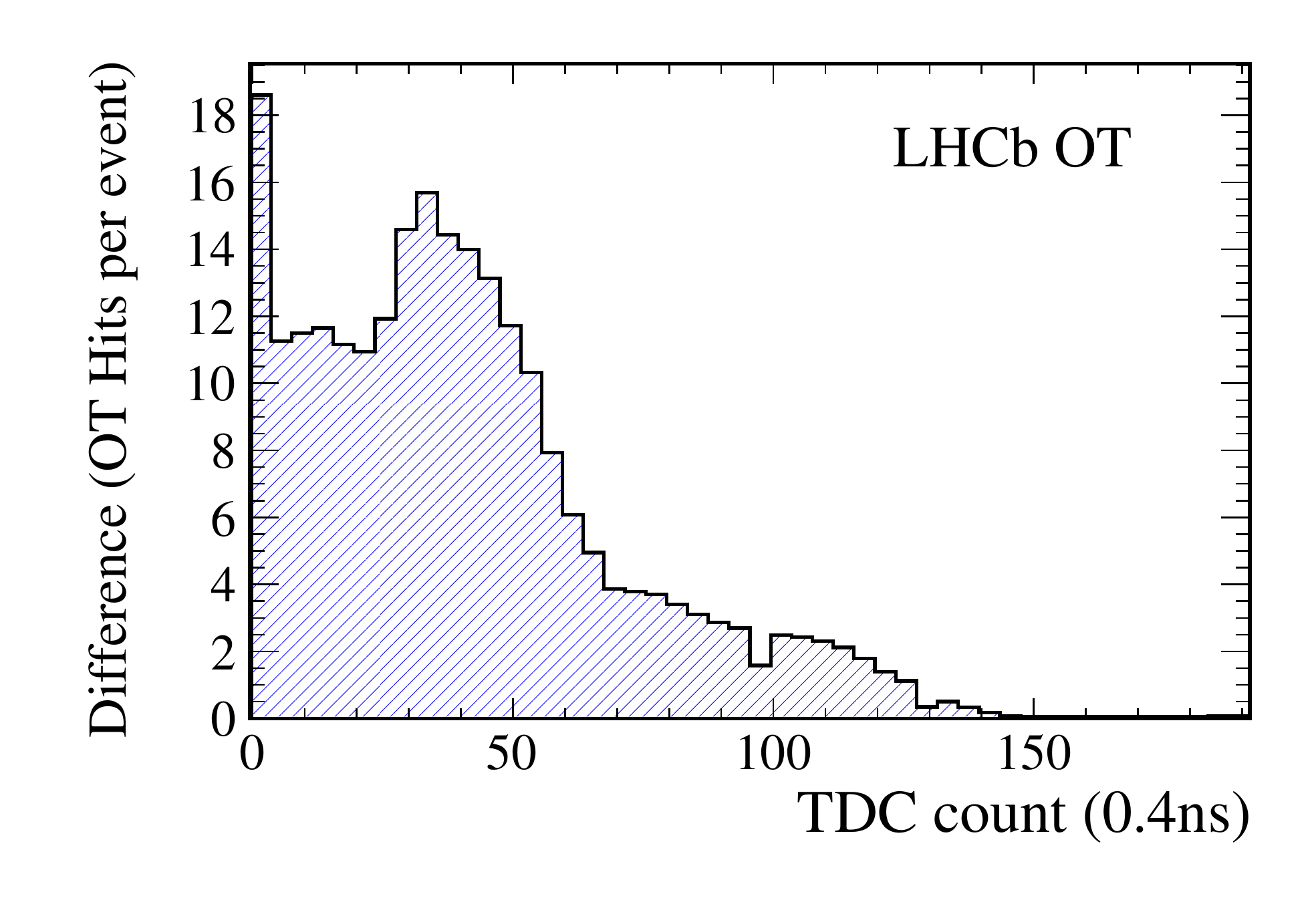}\put(-75,110){(b)}
  \end{center}
  \caption{
    \small The recorded drift time spectrum for all hits in the Outer Tracker
    for no-bias events with $25\ns$ bunch spacing. (a) in blue, the same
    spectrum recorded while keeping only events with $\Sigma E_T (Prev) \leq
    1000$. (b) the difference of the two TDC spectra, scaled such the
    difference is 0 at the $TDC=150$ bin.}
  \label{fig:performance_ratio_of_drfit_times-sumET}
\end{figure}

This feature will be used in 2017 by selecting events with $\sumetprev < 1000 (= 24 \gev)$,
which rejects 7.2\% of the events (figure~\ref{fig:performance-sumET}(b)). 
This does not bias the topology of the physical event
because the event multiplicity in the calorimeter is not correlated between consecutive bunch crossings.
The effect on the drift time spectrum is illustrated in figure~\ref{fig:performance_ratio_of_drfit_times-sumET}(a).
As expected the largest effect of this cut is on the number of (fake) early hits, which contain a large fraction of 
spill-over hits from the previous bunch crossing. 
This is shown in figure~\ref{fig:performance_ratio_of_drfit_times-sumET}(b)
where the difference in TDC spectra before and after the cut is shown, 
normalized to the spectrum at TDC=150.

In addition to the early spill-over hits, also late hits from the current event are rejected, because the whole
event is vetoed. These events can statistically be recovered
with a looser threshold on the transverse energy of L0 hadron clusters.
Effectively, the cut of events with a busy previous event will reduce the average OT event occupancy, 
less affected by spill-over contamination, 
allowing to increase the data-taking rate and speeding up the reconstruction processing time. 

\subsection{Hit asymmetries for both magnet polarities}
The LHCb experiment is specialised in the study of \CP asymmetries, and
a crucial ingredient in these analyses are asymmetries in the detection of charged particles.
To limit the impact from possible detector biases, the magnet polarity
is regularly flipped during data taking.

\begin{figure}[!ht]
  \begin{center}
    \includegraphics[width=0.48\linewidth]{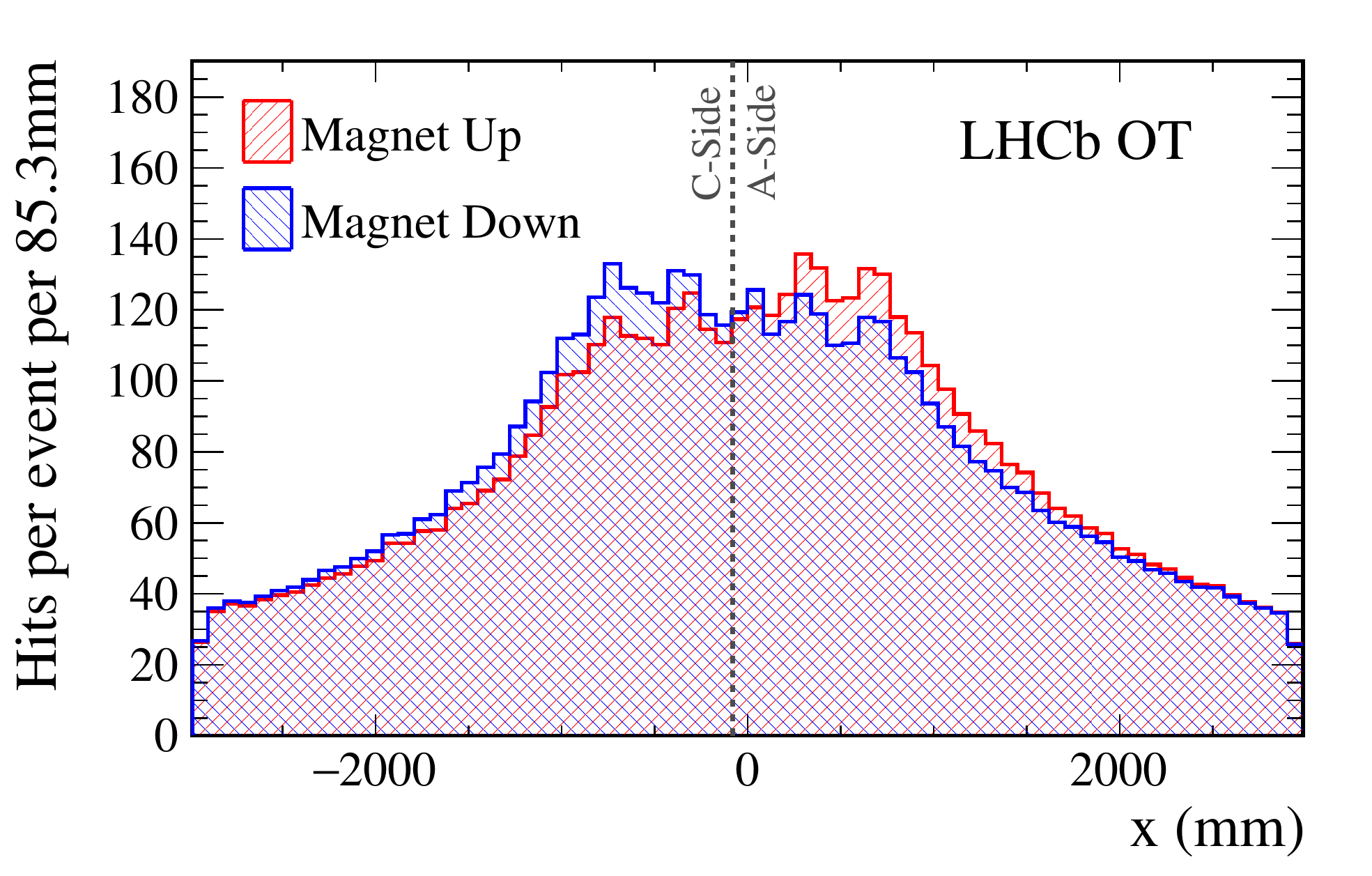}
  \end{center}
  \caption{
    \small The hit distribution is shown for both magnet polarities, 
    using 30,000 random events for both samples, without any trigger requirement.
    The vertical dashed line indicates
    the mechanical split between the two detector halves, at $x=-8$~cm.}
  \label{fig:asymmetry}
\end{figure}

From simulations it is known that more electrons than positrons are produced in secondary 
interactions with the detector material. The electrons are deflected in the so-called ``down'' polarity 
towards the negative $x$-direction (which corresponds to the so-called C-side in LHCb),
and vice-versa for ``up'' polarity. This explains the asymmetric hit distribution in figure~\ref{fig:asymmetry}.

The two detector halves are not of equal width, and the split between the two halves is at $x=-8$~cm. 
As a result, the smaller width of the C-side
registers less hits, which is compensated by the hits from the deflected negatively charged particles
in the ``down'' polarity, leading to only 1.6\% less hits in the C-side than in the A-side.
For the ``up'' polarity, the A-side registers 14.7\% more hits than the C-side.
The total number of hits, averaged for the two polarities, is equal for $x<0$ compared to $x>0$ within $0.4\%$.

\subsection{Hit multiplicities in lead-lead and proton-lead collisions}
The LHCb experiment has expanded its heavy-ion physics programme in Run 2,
by recording \PbPb collisions data in December 2015, and \pPb and \Pbp collisions data in November 2016.
The relatively large OT straw diameter of $4.9\mm$ leads to a 
very large straw occupancy in \PbPb collisions, preventing to reconstruct tracks 
of  \PbPb collisions at the largest centrality.
The offline analysis is limited by the high occupancy in the whole LHCb detector 
 to an event centrality smaller than $60\%$.

\begin{figure}[h!]
  \begin{center}
    \includegraphics[width=0.48\linewidth]{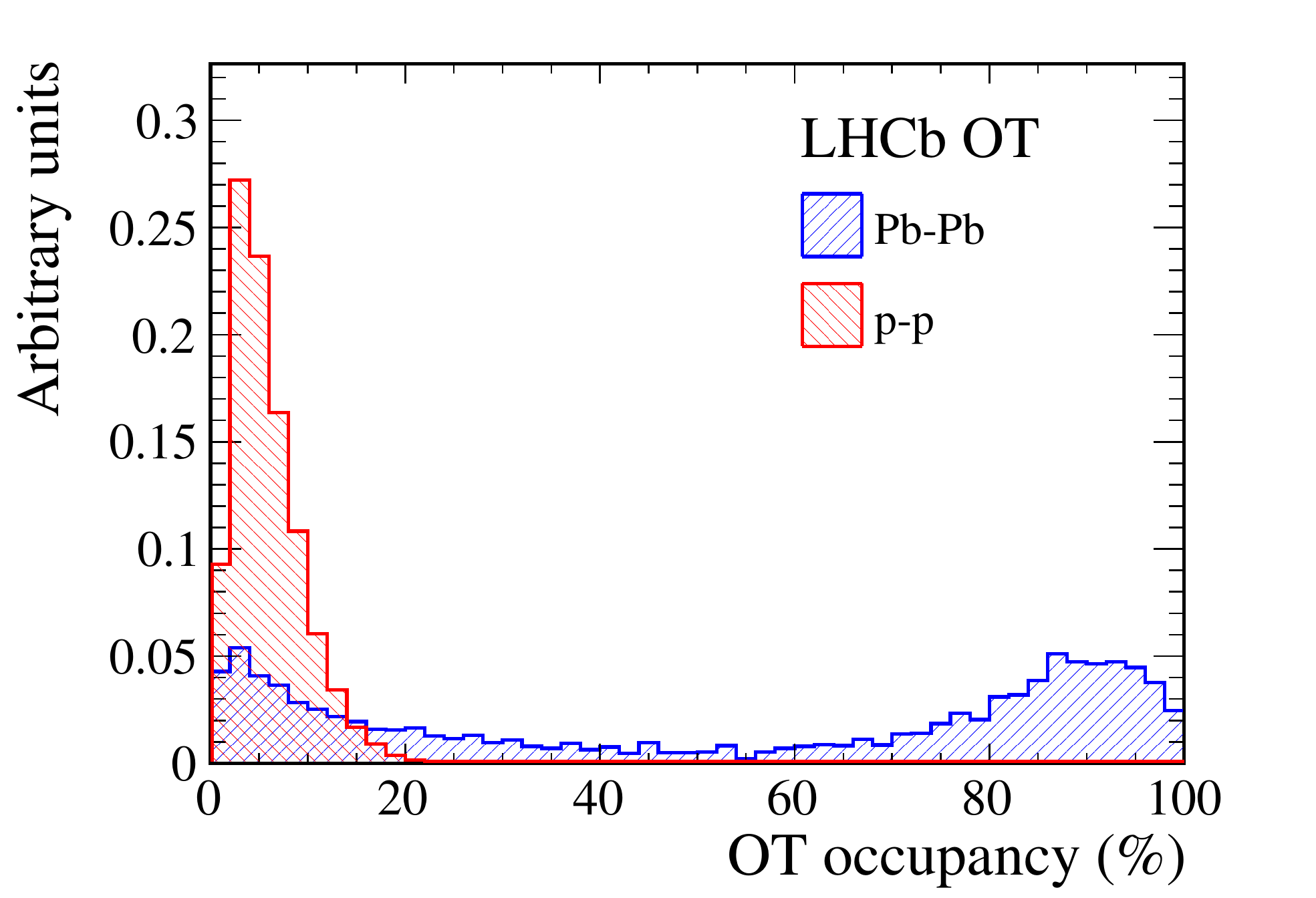}\put(-160,120){(a)}
    \includegraphics[width=0.48\linewidth]{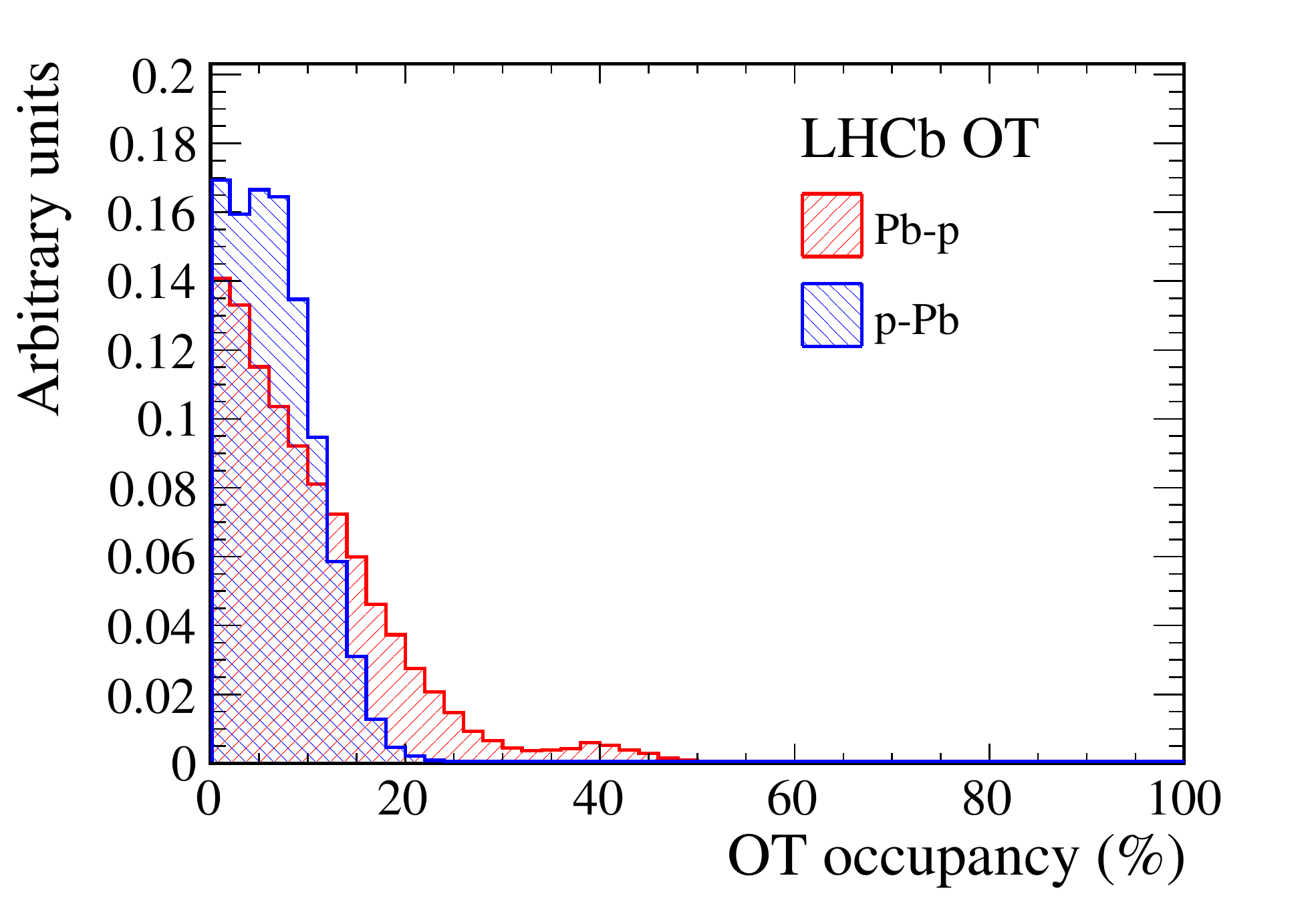}\put(-160,120){(b)}
  \end{center}
  \caption{
    \small The average number of recorded hits (a) in minimum bias proton-proton and \PbPb collisions
 and (b) \pPb and \Pbp collisions.}
  \label{fig:performance-Pb}
\end{figure}

The OT occupancy for \PbPb events is shown in figure~\ref{fig:performance-Pb},
compared to the occupancy in nominal $pp$ collisions, with an average pile-up of $\mu=1.1$.
The occupancy in $p$Pb and Pb$p$ collisions is also given, showing a slightly larger occupancy
for the situation where the lead-ions are circulating in beam-1, i.e. in the direction of the LHCb 
spectrometer.

\clearpage 
\section{Mechanical stability}
\label{sec:rasnik}

An alignment system was constructed to monitor the relative movements of 
the OT detector frames. The main purpose of the system is to confirm the mechanical 
stability of the detector. Any unforeseen movements of the detector components are
monitored and can be used to confirm, or used as constraint for, the alignment procedure with reconstructed
charged particle tracks~\cite{Hulsbergen:2008yv, Amoraal:2012qn}.

The system used for this purpose is called the Relative Alignment 
System of NIKHEF (RASNIK)~\cite{Duinker:1988se,*Rasnik-2001-004,*Rasnik-2002-016}.
A finely detailed image is projected through a lens onto a CCD camera. 
The light source is a $3\times 3$ grid of infrared-emitting LEDs. 
These LEDs illuminate a coded mask. The mask consists of black-and-white 
squares in a chessboard-like pattern. 
The coded non-repeating pattern is used to obtain a unique position. 
The movement of the lens by a distance $d$ in a direction perpendicular 
to the axis defined by the mask and the CCD camera causes a displacement 
of the image of the mask by a distance $2d$. The transverse position 
of the lens is then calculated from the image position by means 
of image processing of a CCD pixel frame. 
The image recorded by the CCD camera is then used to
reconstruct positions of the chessboard pattern. 
The resolution of the CCD-RASNIK system  
is better than $\rm 1 \mum$ for a system with 
a mask-to-CCD distance up to about $8\m$~\cite{Dekker:1993qq}. 

Elements of 48 RASNIK lines are mounted on the four corners
of each of the 12 C-frames of the OT detector. They measure precisely the $x$ and $y$ displacements of the four 
points on the C-frame with respect to corresponding fixed reference points,
mounted on the bridge, the pillars or on the table fixed to the support structure. 
In addition to these 48 short horizontal lines, there are two longer vertical lines 
measuring displacements of the $x$ and $z$ coordinates of two positions
on the top of the support bridge, with respect to the supporting table.

Below, the measured displacements are discussed regarding the long term stability,
the reproducibility of the position after opening and closing a C-frame and  
the movements due changes of the magnet field of the LHCb dipole.
All the misalignments here presented are automatically taken into account and corrected 
for via the LHCb software alignment~\cite{BORGHI2017560}.

\subsection{Long term stability of C-frames positions}

\begin{figure}[!t]
\begin{center}
\begin{overpic}[width=0.49\linewidth]{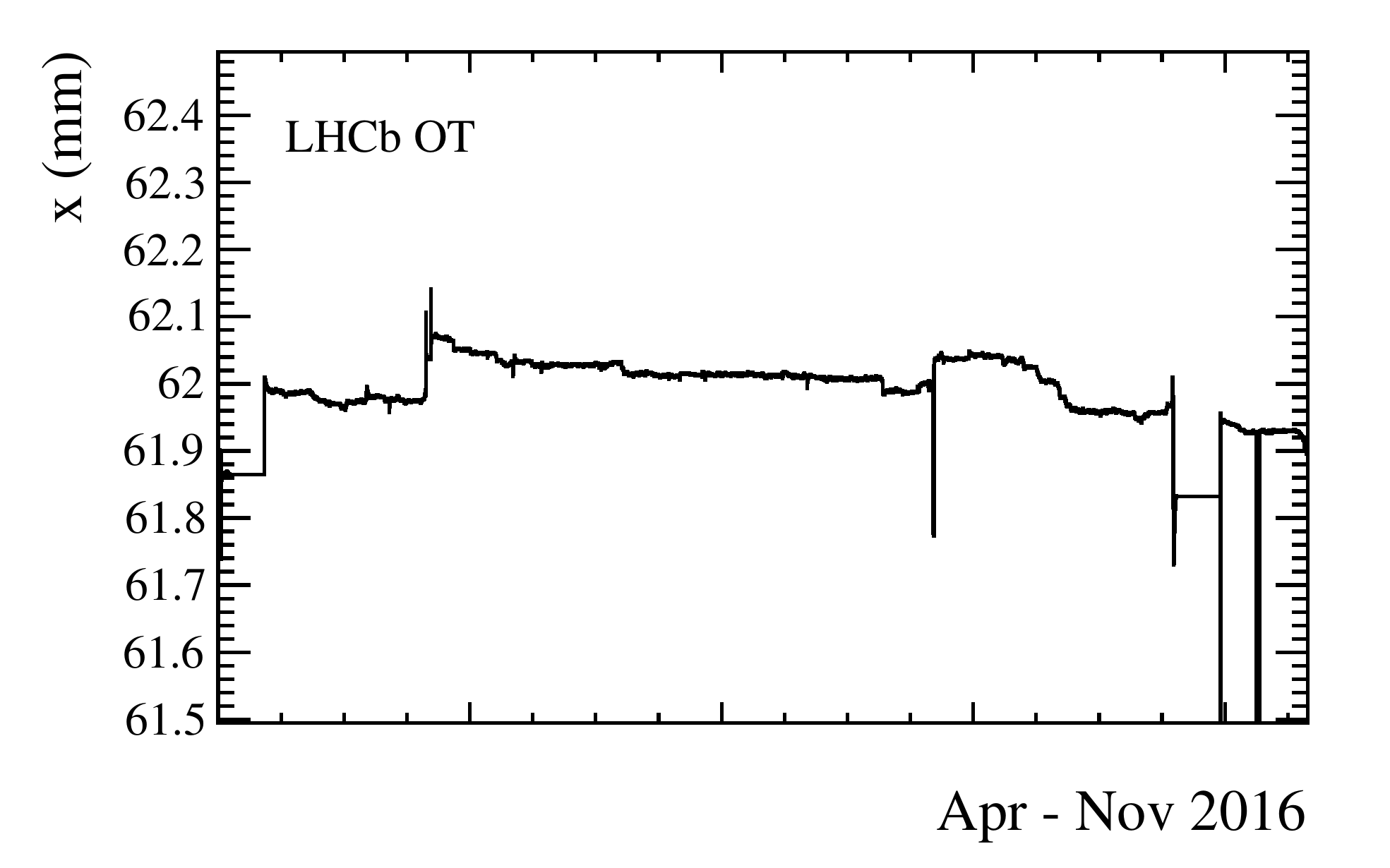}
\put(77,50){\small (a) }
\end{overpic}
\begin{overpic}[width=0.49\linewidth]{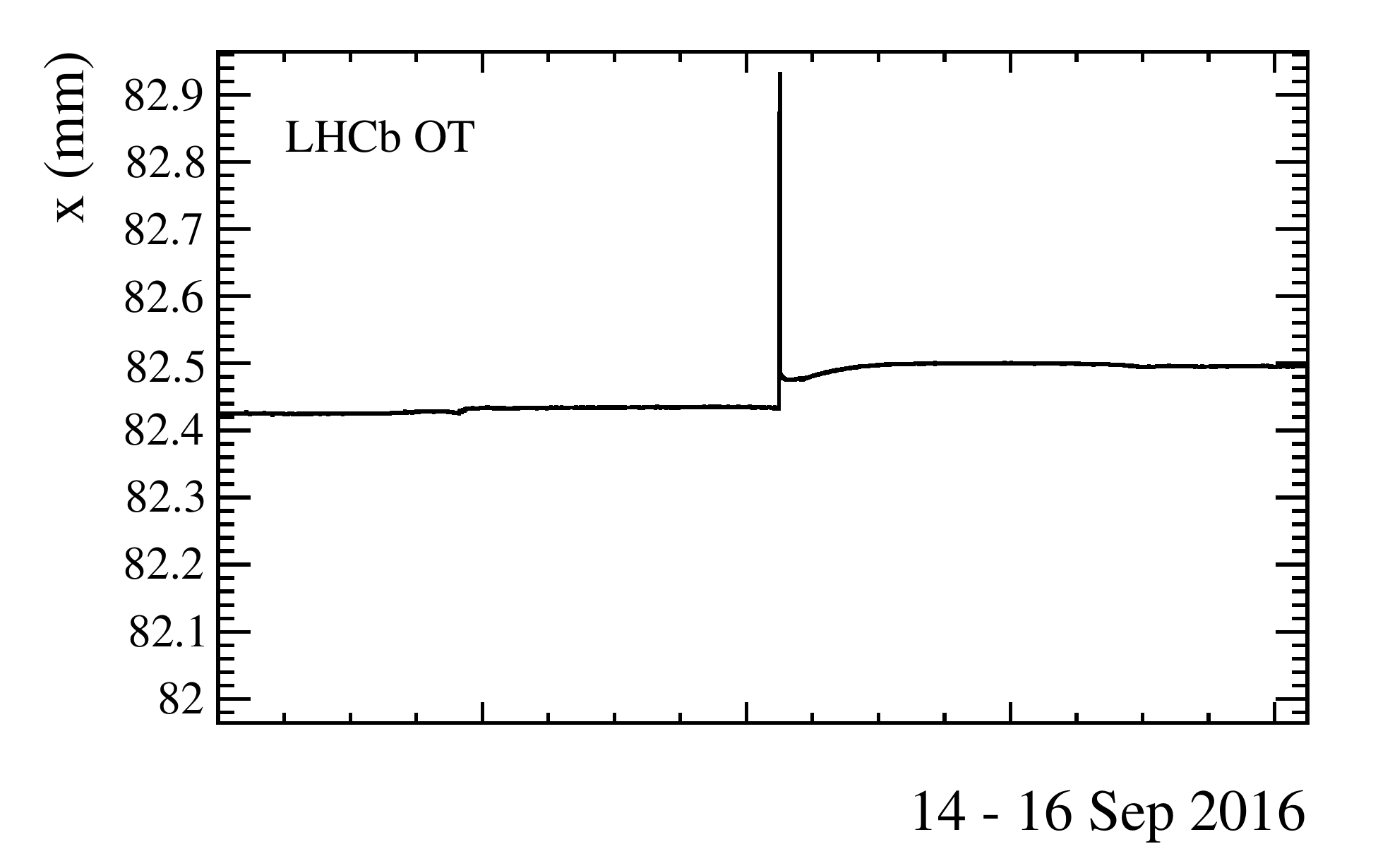}
\put(77,50){\small (b) }
\end{overpic}
\caption{\small (a) Movements in $x$ of one bottom corner of a C-frame (T2-XU) 
              are shown during the data taking period in 2016.
(b) Movement in $x$ of a point at the top corner
          of an opened and closed C-frame (T3-VX, C-side) in September 2016. 
(The absolute value of the $x$-coordinate is arbitrary.)}
\label{fig:T2XUFullYear}
\end{center}
\end{figure}

The horizontal lines measure the horizontal $x$ and  vertical $y$ variations
of positions of the points close to C-frames' corners.
In figure~\ref{fig:T2XUFullYear}(a) the movements along the $x$ coordinate are shown, which is 
the coordinate relevant for the momentum measurement of charged particles.
The RASNIK results show that the top parts of the C-frames are more
stable in time than the bottom ones in both $x$ and $y$ coordinates.  This is
consistent with the mechanical constraints of the C-frames, which are fixed in
$x$ from the top of the bridge by means of a hook, clamped to the C-frames.  The
bottom of the C-frames are hanging freely in the $x$ and $y$ coordinates, while
constrained in the $z$ coordinate.
The positions of the bottom C-frames vary within $\sim 200\mum$ in both $x$ and $y$ values in 2016. 
At the beginning of the data taking, in May and June 2016,
the changes are relatively large, in the range $100-200\mum$. 
After an intervention at the end of June when parts of the detector were opened and closed,
the OT slowly attained the equilibrium state. 
After a second intervention in September the OT reached its equilibrium in a shorter time.  
In October and November, the changes start to evolve 
with the opposite trend to the ones observed in May. 

\subsection{Effects of opening and closing C-frames}

Two C-frames ($x$-$u$ and $v$-$x$ of the 3$^{rd}$ station -- T3-XU and T3-VX -- on the C-side) 
were opened and closed on $15^{\rm th}$ September 2016. The data from
one day before and one day after the movement are presented in figure~\ref{fig:T2XUFullYear}(b). 
The horizontal shift of about $70 \mum$ in the $x$ is observed due to the intervention.
The vertical $y$ coordinate changed by only $20\mum$. It was also checked that 
the intervention on station T3 resulted in significantly smaller movements, within $\pm 20 \mum$,
of the other two stations that were not moved during the intervention.
Changes in the weight distributions on the bridge 
during opening and closing C-frames, slightly affect all C-frames due to changing weight distributions 
of the bridge.

\subsection{Monitoring of the bridge position}

\begin{figure}[!h]
\begin{center}
\begin{overpic}[width=0.49\linewidth]{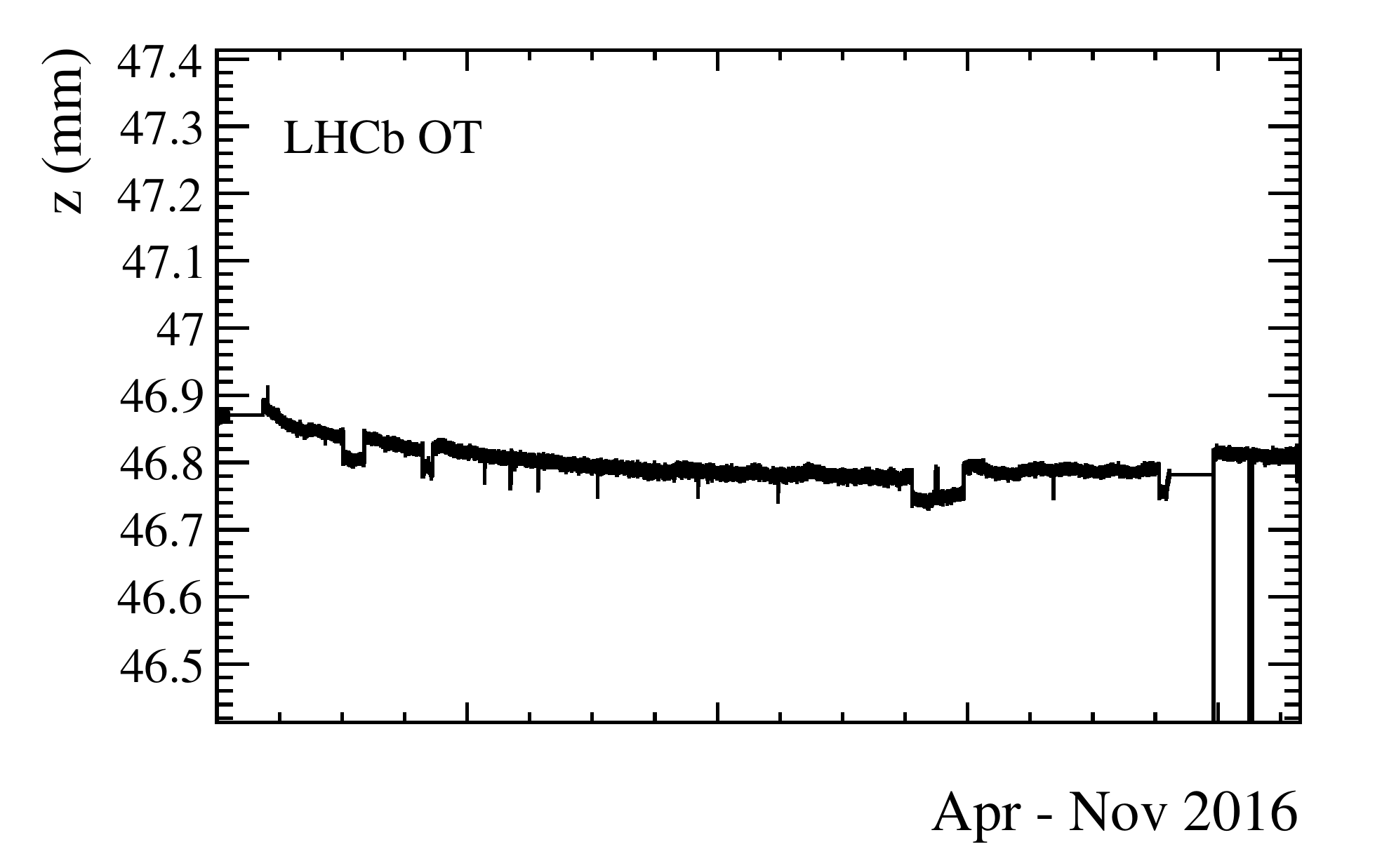}
 \put(77,50){\small (a) }
\end{overpic}\begin{overpic}[width=0.48\linewidth]{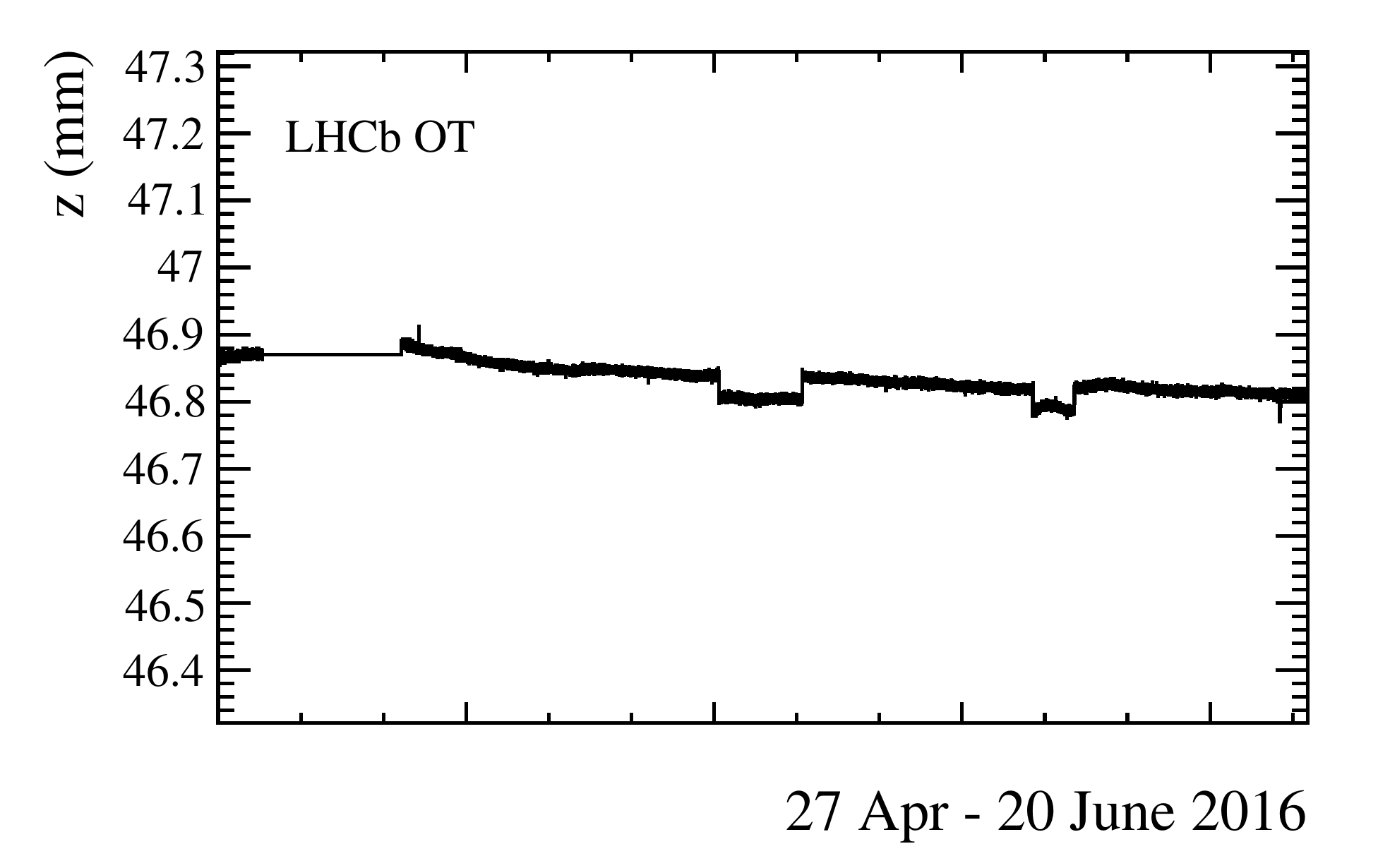}
  \put(77,50){\small (b) }
\end{overpic}
\end{center}
\caption{\small The $z$--coordinate of the bridge positions as a function
              of time (a) for the data acquired in 2016  and (b)
              the more detailed view of the data from 27th of April 
              till 20th of June. The two visible dips correspond to the two occasions where the
      magnet was switched off and on. }
\label{fig:BridgePosition}
\end{figure}

There are two lines which allow to measure the $x$ and $z$ LHCb coordinates
of the points close to the top and on either side of the supporting bridge.
The long term movement in $z$ is shown in figure~\ref{fig:BridgePosition}(a), 
whereas a zoom is shown in figure~\ref{fig:BridgePosition}(b)
in the period when the magnet was switched on and off several times. 
In May and June 2016, the bridge moves towards the magnet by about $100 \mum$ and
stabilizes in August and September. There is a hint that the trend reverses 
towards the end of the year and the $z$ starts to increase. 
The sudden movements observed correspond to periods where the LHCb dipole magnet 
was switched off and on.  The long term movement of the order $100\mum$ in the $z$ direction
is small, and is presumably due to the combination of temperature variation and magnetic forces.

The switching off of the magnet causes the movements of 
the bridge. After switching on the magnet, the bridge moves
back to the previous position. The bridge position does not depend on the
direction of magnet polarity.

In conclusion, the RASNIK system works well and provides 
high accuracy data which allows to monitor deformations of the OT detector geometry.
The movements of the C-frames in the $x$ and $y$
coordinates are smaller than $200\mum$.
The movements of the bridge in the $z$ direction with respect to the supporting table are within
 $100\mum$.
The results show small movements of the OT due to changes of
magnetic field configurations and mechanical interventions, like opening and closing of C-frames.
The RASNIK lines confirm the software alignment with precision
data and show the real movements of the detector.

\section{Radiation resistance}
\label{sec:ageing}
During the production of the OT detector modules, significant gain losses were
measured after irradiation with $^{90}$Sr or $^{55}$Fe sources. This effect was studied extensively,
and the origin was traced to the plastifier di-isopropyl-naphthalene used in the glue,
which forms an insulating layer on the anode wire~\cite{Bachmann:2010zz,Tuning:2011zzb}. After
local irradiation, surprisingly the most affected region is upstream of the
source with respect to the gas flow. Downstream the source, the deposition of the
insulating layer is prevented by the presence of ozone, produced under the source.
The irradiation profile on the OT surface during LHC operation is very different
from the one of a small source in the laboratory, resulting in the fact that the entire OT 
benefits from the ozone production, and thus does not suffer from malicious depositions
of an insulating layer of hydro-carbons on the anode wire.
Nevertheless the gain is monitored regularly during operation, 
to detect possible irradiation damage at the earliest stage. 

To determine any change of the gain, dedicated data are recorded
during LHC operation in which the amplifier threshold is varied per layer. 
To limit the impact on physics data taking, typically 
the low intensity fills from the LHC are used.
However, the LHC conditions during these OT threshold scans can vary.
These varying beam conditions lead to different event occupancies in the Outer Tracker.
This was traced to be the origin of the small reported increase of the gain in the Outer Tracker
in 2012, which was falsely attributed to a change in the gas mixture~\cite{vanEijk:2012dx}.

\begin{figure}[!h]
  \begin{center}
     \begin{overpic}[width = 0.49\textwidth]{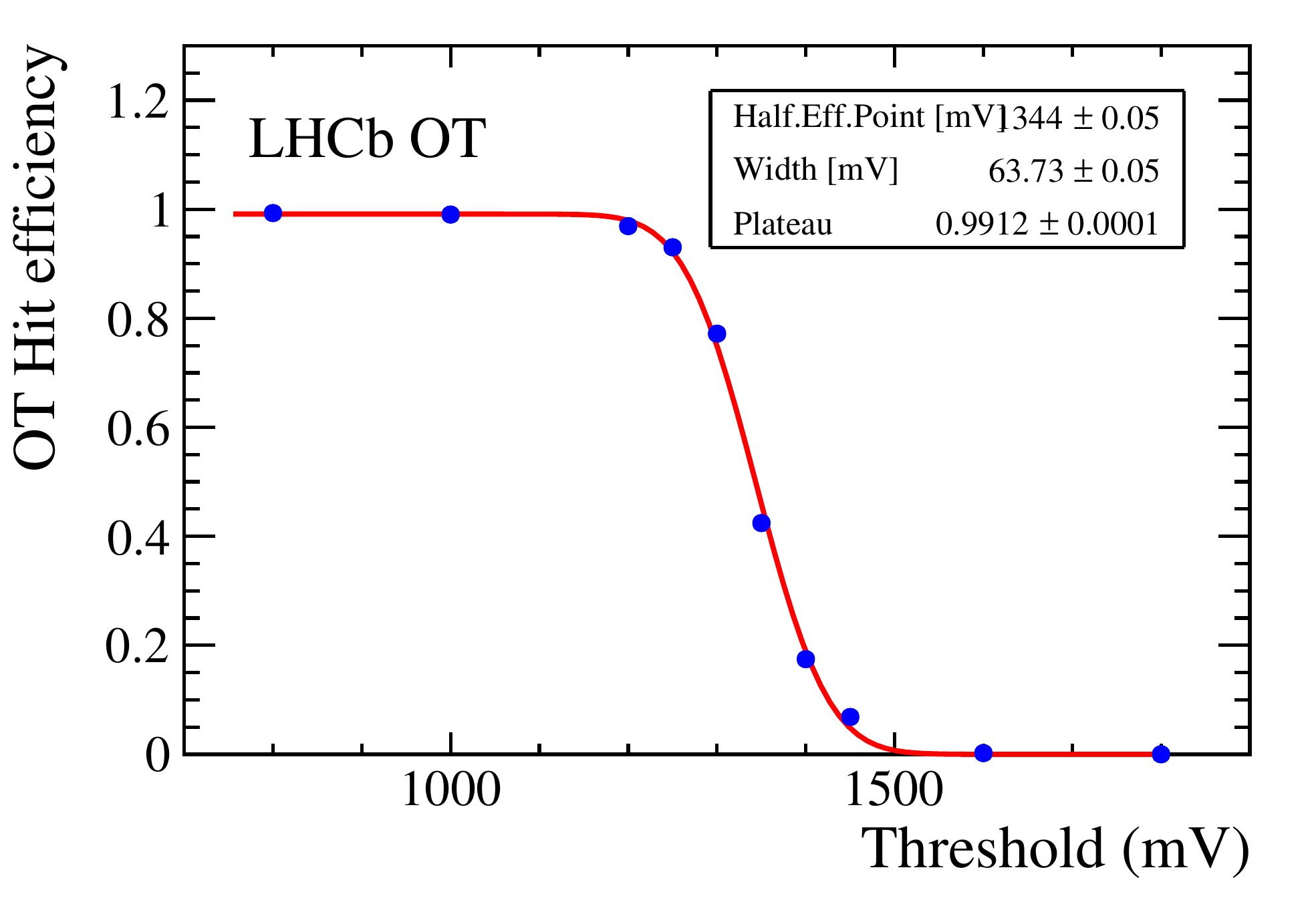}
       \put(26,45){(a)}
      \end{overpic}
      \begin{overpic}[width = 0.49\textwidth]{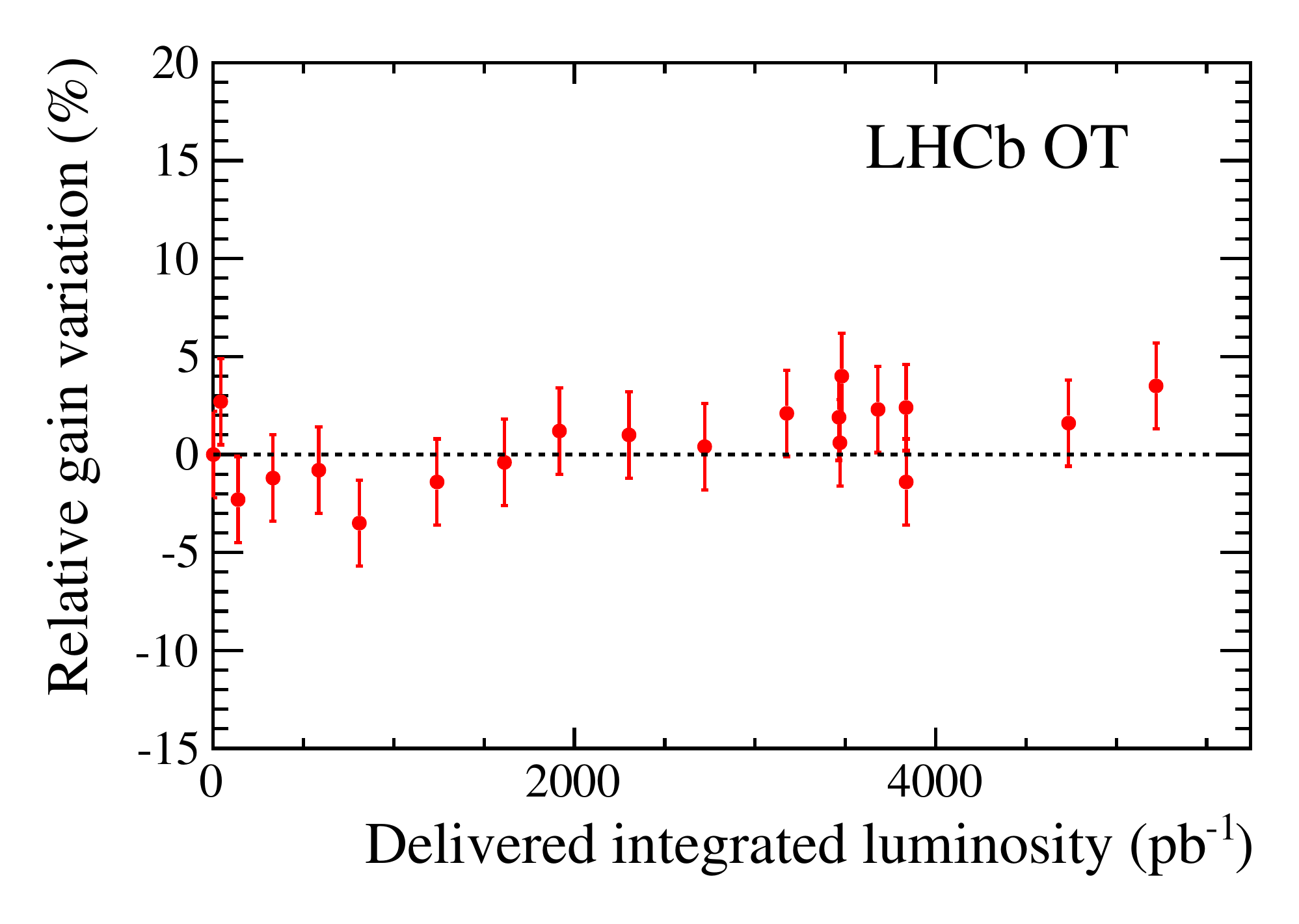}
       \put(65,50){(b)}
      \end{overpic}
     \end{center}
  \caption{
    \small   (a) Example hit efficiency plot as a function of amplifier threshold 
for the first layer. (b) Average gain variation as a function of the delivered luminosity. }
  \label{fig:ageing_trendplot}
\end{figure}

The measured average gas gain with the nominal operational conditions is 55\,000 and the charge deposited per single hit
is of 350~fC. The typical hit density corresponds to about $2.8 \times 10^{-3}$  per event per \cm per straw
in the innermost part, which is the one with highest multiplicity.
For the considered time period a delivered luminosity of 5.2\invfb was integrated
for an equivalent charge of $0.18~\rm{C}/\cm$ per straw. 

Data from all the threshold scans have been re-analysed.
To equalise the occupancy for each scan, an upper limit on the number of OT hits per event
is placed offline. The single hit efficiency is
computed in bins of the threshold value, which gives rise to the characteristic
S-shape as shown in figure~\ref{fig:ageing_trendplot}(a). The relative
gain is then computed from the change in the half-efficiency point~\cite{vanEijk:2012dx}.
The relative gain throughout the years averaged over the full detector is shown
in figure~\ref{fig:ageing_trendplot}(b) as a function of the integrated
luminosity. The results are consistent with no gain loss. The inner, outer and
bottom parts of the detector are analysed separately as well, motivated by
the originally observed ageing pattern upstream of the irradiated area. No differences
are observed for the different regions of the detector.
Within the resolution of the method (approximately $34\times34 \cm^2$) no large gain variations are seen 
in localized parts of the detector or in single straws.

\cleardoublepage
\part{Improved calibration and results}

In this second part the improvements implemented during the long-shutdown
are presented and evaluated. 
Prospects for future use of the added information are discussed.

\section{Real-time global drift time calibration}
\label{sec:realtime}

The monitoring of the precise time alignment of the LHCb detector with respect
to the LHC clock partially 
relies on the OT detector. The calibration of the time-alignment was performed
in Run 1 
only at given time intervals, typically after periods of shutdown with hardware
interventions (technical stops).
In order to increase the precision and automation of the OT time-alignment 
a procedure has been devised to perform this task online during data-taking time,
and is described in the following. 

\begin{wrapfigure}{R}{0.39\textwidth}
  \begin{center}
    \includegraphics[width=1.1\linewidth]{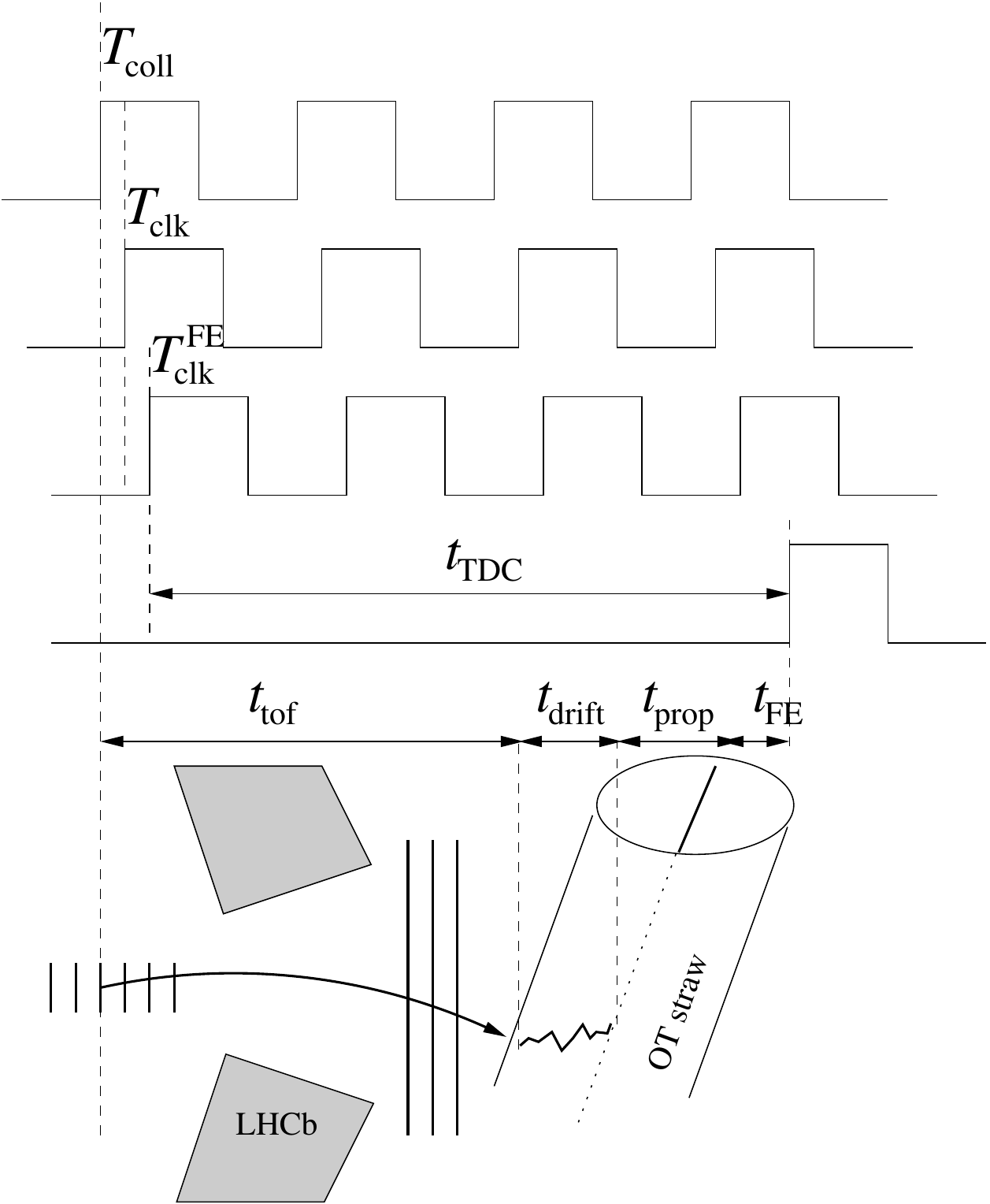}
  \end{center}
  \caption{
     Sketch of the contributions to the measured $t_{TDC}$.}
  \label{fig:t0-sketch}
\end{wrapfigure}

\subsection{Method}

The OT measures the arrival time of the signals with respect to
the LHC clock, 
$T_{\rm clk}$, and is referred to as the TDC time, $t_{\rm TDC}$. 
From the measurement of the distance $r$ of a track traversing the straw to the 
wire at the centre of the straw, 
the expected time, $t(r)$ can be calculated using the distance to drift-time relation 
(TR-relation, see also section~\ref{sec:rt-relation}).

Different contributions can be distinguished in the TDC timing: 
the time-of-flight of the particle, $t_{\rm tof}$, the drift-time $t_{\rm
drift}$ of the electrons in the straw,
 the propagation time of the signal along the wire to the readout electronics,
$t_{\rm prop}$, and the delay induced by the FE electronics, $t_{\rm FE}$, as shown in figure~\ref{fig:t0-sketch}.

\begin{equation}
t_{\rm TDC} = (T_{\rm collision} - T_{\rm clk}^{\rm FE}) + t_{\rm tof}+t_{\rm drift} + t_{\rm prop} + t_{\rm FE}
\end{equation}

The bunch crossing time of the $pp$ collision, $T_{\rm collision}$, can be shifted 
with respect to the LHC clock,  $T_{\rm clk}$, but the difference is  kept below $0.5\ns$.
The phase of the clock used by the OT electronics at the TDC input ($T_{\rm clk}^{FE}$) can be 
adjusted.
The sum $ (T_{\rm collision} - T_{\rm clk}^{\rm FE}) + t_{\rm FE}$ can be
written as ${(T_{\rm collision} - T_{\rm clk}) + t_{0}}$, where the first difference
accounts for variations of the phase of the LHC clock received at the LHCb experiment, and the 
$t_{0} = t_{\rm FE} -t^{\rm FE}_{\rm clk}$ offset accounts for the drift of
the global LHCb clock and the drift of FE electronic delays. 

The TR-relation is calibrated on data (see section~\ref{sec:rt-relation}) and the time $t(r)$
 obtained includes the effects of $ t_{\rm tof}$, $t_{\rm drift}$ and $t_{\rm
prop} $. 
The measured difference between $t(r)$ and $t_{\rm TDC}$ is fitted with a Gaussian distributions, 
which mean value determines the offset $t_{0}$.

\begin{figure}[!t]
  \begin{center}
    \includegraphics[width=0.475\linewidth]{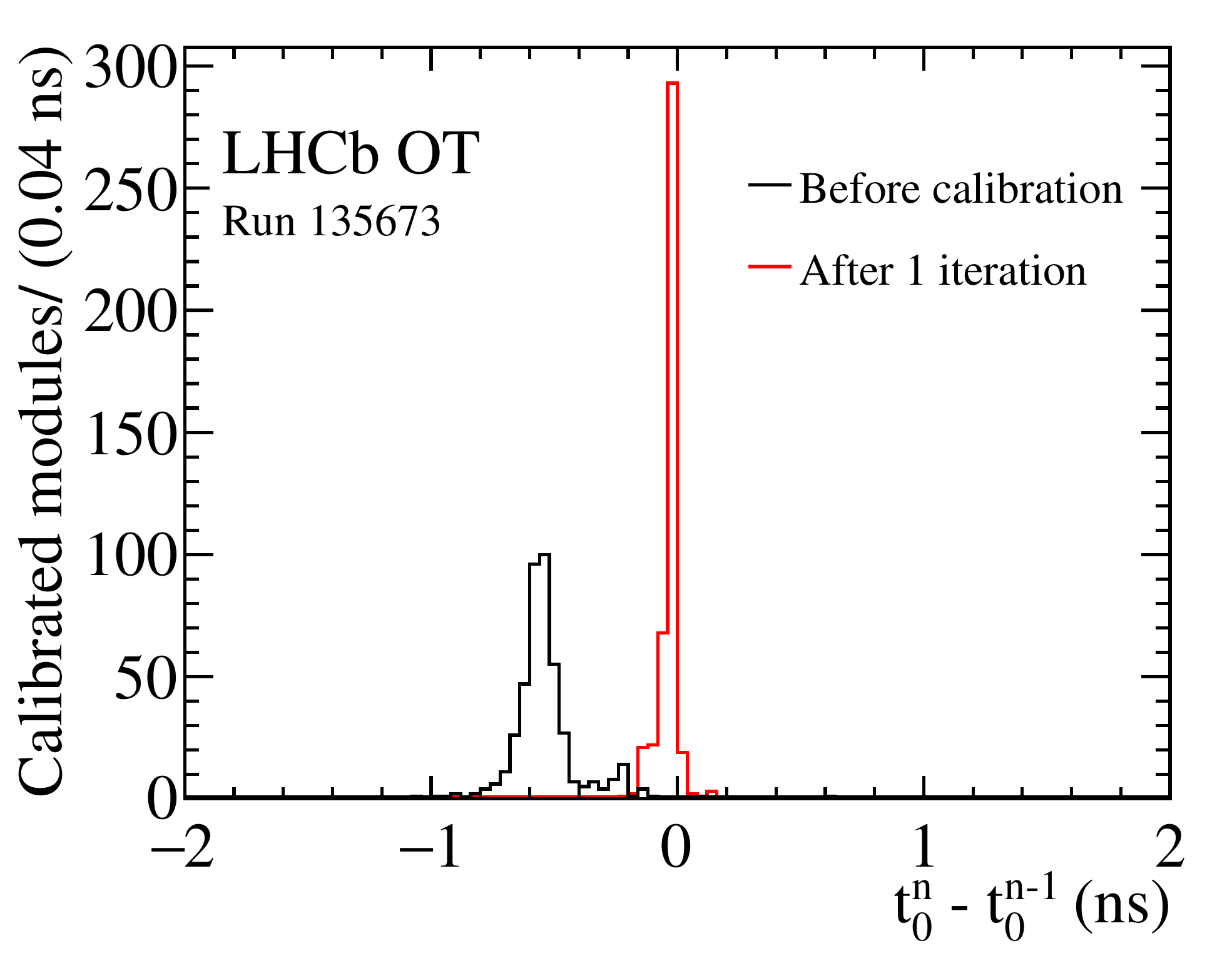}
 \vspace*{-0.5cm}
  \end{center}
  \caption{
    \small
    Example distribution of the differences between the $t_{0}$ values before and after the first iteration of the calibration (black), 
and between the first and the second iteration (red).}
  \label{fig:1}
\end{figure}

During Run 1, the $t_{0}$ calibration was executed after
every LHCb Technical Stop and the $t_{0}$ offset for each OT module was determined from 
the drift time residual distributions~\cite{Arink:1629476}.
One iteration of the calibration procedure was sufficient to obtain the new  $t_{0}$ values with a
precision better than $0.1\ns$. 
The distributions of the differences between the $t_{0}$ values before calibration and after the first iteration, 
and between the first and the second iteration are shown in figure~\ref{fig:1}.
The largest contribution
is due to a global $t_{0}$ shift that affects the $t_0$ constants of all FE modules by the same amount.
This is caused by the drift of the global LHCb clock with respect to the bunch arrival time. 

During Run 2, the $t_{0}$ calibration procedure is therefore split in two algorithms. 
The real-time algorithm performs a gaussian fit to the global drift time residual distribution
considering the tracks in the whole Outer Tracker, estimating the impact of the
drift of the LHCb clock. In addition, an offline algorithm is set up
to evaluate the time offsets per OT module due to the single FE electronic delays. These
latter offsets are not expected to change during the data-taking period,
excluding cases of hardware interventions on the OT modules.
This second algorithm has been used to calculate the offsets at the beginning of
Run 2 and is then only used to confirm the stability of the $t_0$ constants (see section~\ref{sec:t0module}).

\subsection{Performance and monitoring}

\begin{figure}[!b]
  \begin{center}
    \includegraphics[width=0.9\linewidth]{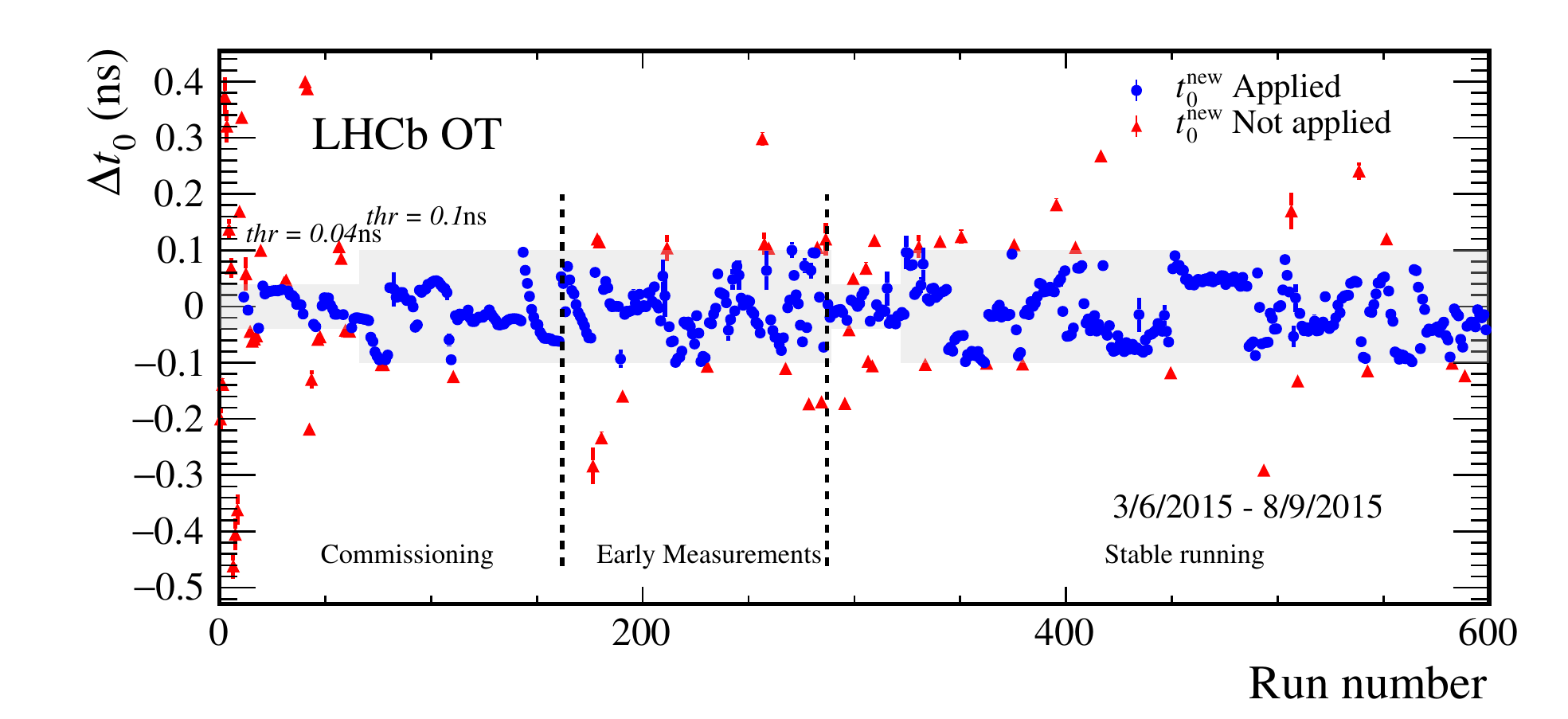} %
 \vspace*{-0.5cm}
  \end{center}
  \caption{
    Calculated global drift-time offset differences ($\Delta t_{0}$) for Run 2
data taking period 3 July 2015 - 14 Sept 2015. 
Red triangles show calibrations above threshold which are applied to data, 
while blue dots show calibrations which are within tolerance and are not applied.
 The threshold values used are represented with the shaded region: 
the periods with lower threshold were during commissioning of this algorithm. }
  \label{fig:2}
\end{figure}

The real-time global $t_{0}$ calibration strategy allows for a time alignment of
the OT time and LHCb clock better than $0.1 \ns$, compared to global shifts of $\pm 0.5 \ns$ 
which were allowed in Run 1. 
This leads to an increase of the
track reconstruction efficiency by 0.25\% with respect to the previous calibration strategy. The track
reconstruction efficiency has been estimated from
a sample of simulated $\PB^{+}\to \jpsi K^{+}$ decays reconstructed in the LHCb
detector. The efficiency of tracks reconstructed in the whole LHCb tracking
system, using the $t_{0}$ values included in the
LHCb detector simulation, is compared to the tracking efficiency measured
when varying the simulation $t_{0}$ offset by a global shift. 

The real-time calibration algorithm uses the drift time residuals distributions
obtained for the tracks reconstructed in the three OT stations to measure the
current drift time offset ($t_{0}$ value) with respect to the $t_{0}$ value
calculated from the previous calibration. This calibration is used in the online
database during data taking. This calibration algorithm produces a fit to
the global drift time residuals distribution every 15 minutes during physics data taking.
The new $t_{0}$ value is applied if the following conditions 
are met: sufficient tracks were used in the new calibration; no $t_{0}$ calibration 
had been performed yet on the current run; the new $t_{0}$ value change exceeds 0.1~ns 
but is smaller than 2~ns. 

The difference between the new $t_{0}$ value with respect to the previous $t_{0}$ value 
is reported in figure~\ref{fig:2}.
The shadowed regions correspond to the minimum size of this difference necessary
to trigger an update of the $t_{0}$ constants in the databases used for the
data taking.
The threshold value of  $0.1 \ns$ is considered optimal.
It is the typical spread of the $t_{0}$ values per
module after one or more calibration iterations, and thus implies a natural choice for the intrinsic resolution. 
Typically a few updates of the global $t_{0}$ constant are applied in one week of data taking. 

\begin{figure}[!t]
  \centering 
      \includegraphics[width=0.99\linewidth]{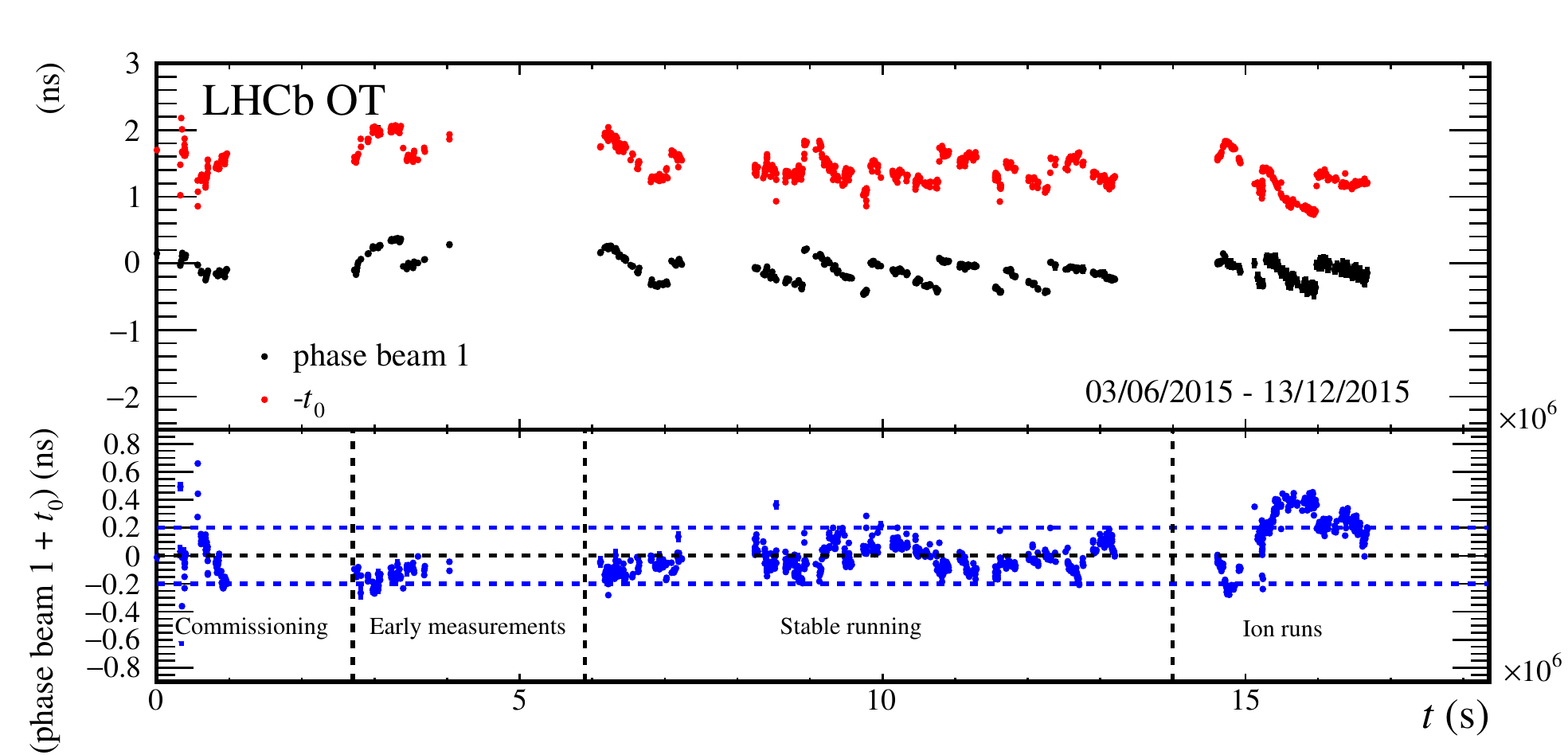}
  \caption{
    Absolute stability of the value of the global time offset $t_0$
obtained comparing the $t_0$ absolute value with the beam-1 phase.
The large scatter in the initial points is due to the phase of commissioning of the algorithm. 
The $t_0$ measurements towards the end of the 2015 data taking period correspond to the LHC operation with Pb ions.  }
    \label{fig:absolute_t0}
\end{figure}

The absolute $t_0$ as a function of time is shown in figure~\ref{fig:absolute_t0}, 
together with the absolute value of the phase of beam-1. Any change in the LHC clock affects both measurements.
The difference of the two measurements is shown in the lower figure of figure~\ref{fig:absolute_t0},
and is stable within $\pm 0.2 \ns$, showing the stability of the average OT time measurement.
The large scatter in the initial points is due to the phase of commissioning of the algorithm. 
The larger differences $> 0.2$~ns in the last data taking period correspond to the data recorded
during Pb running period at the end of 2015. Due to the larger number of low momentum tracks, 
the average time-of-flight difference is reflected in a different global $t_0$.

\section{Offline per-module calibration}
\label{sec:singlefe}

Time offsets due to small differences in the delays in single Front-End modules 
need to be accounted
for, in addition to the global $t_{0}$ offset. These FE-offsets are not expected
to change as function of time, unless in cases of hardware modifications of the
system. Every FE module of the OT contains two layers of 64 straws, read out in groups
of 32 channels, corresponding to one TDC chip per group. Two
changes in Run 2 improve the time calibration of the OT module.

First, the calibration procedure assumes that the spatial position
and alignment of the OT is correct. Since the calibration exploits the tracks
and compares the measured time $t_{\rm TDC}$ to the time calculated from the
measured distance from the track to the wire, $t(r)$, an imperfect spatial alignment
of the detector will result in a systematic overestimation or underestimation of $t(r)$. 
The relative position of the two monolayers of straws inside a single module 
has been determined and accounted for in Run 2.

Secondly, the four different TDC chips within the same module measure different $t_{\rm
TDC}$ values, due to the slightly different characteristics. The $t_{\rm
TDC}$ value measured can vary up to 1~ns within the same module. 
In Run 2 the granularity of the $t_{0}$ constants has been increased by a factor four,
by determining the constant per TDC-chip, rather than per FE-module.

\subsection{Module and monolayer spatial alignment}

A major improvement of the alignment of the detector modules in Run 2 is provided
by the real-time spatial alignment strategy~\cite{BORGHI2017560}. At the beginning of each fill,
a full tracker alignment is performed, and when the constants describing the
position of the module differ from the last constants by more than a given
threshold, the constants are updated. This ensures to have the OT modules aligned
to better than 100\mum at all times.

Not only the alignment constants are updated more frequently, 
also the detector geometry description is improved. 
More parameters are introduced to describe the shape of a detector module.  
The OT half-module geometry is now modelled as a collection of 3
straight segments with equal length~\cite{Arink:1629476}. The numbers that parametrize the
misalignment of the two monolayers are stored in the condition database. There
are at most six numbers for each module. These six numbers represent the
displacement in the $x$-coordinate between the two monolayers at the beginning and
end of each of the three trajectory segments. In order to perform this special
alignment, data from Run-I has been used. This new description of the detector elements
leads to a reduction of the width of the drift time residuals
distributions by up to 20\% (See section~\ref{sec:rt-relation}) 
and a marginal increase of the number of reconstructed tracks and OT hits. 
The monolayer alignment is shown to be stable and is not updated online, 
but is monitored offline during Run~2.

\subsection{FE-module  \texorpdfstring{$t_{0}$}{} calibration}
\label{sec:t0module}

The $t_{0}$ offsets for single TDC chips are calculated  with respect to the global
$t_{0}$ offset. This calibration is only run periodically as no relative movements
of the modules are expected.
For each module the drift time residual distributions are drawn for each TDC chip
separately for tracks traversing the straws from the left and right side. A
gaussian fit excluding the tails of the distributions is performed to determine
the position of the drift time residuals peak. An arithmetic average between the
 $t_{0}$ obtained from left and right tracks is considered, and a $t_{0}$ value
for each constant is determined and reported in the databases. 

Different selection studies have been performed, including different requirements on the minimum
numbers of OT hits and different lower limits on the momentum of the tracks.
A tighter selection ensures a better quality of the track sample,
but reduces the statistics. A compromise between the acceptable statistical
uncertainty and track quality requirements is reached when using tracks
traversing the whole LHCb tracking system, requiring  
an unbiased track-fit $\chi^{2}/\text{ndf}<2$ 
and a minimum momentum of the particle of $3.2 \gevc$, 
which are standard requirements in typical LHCb physics analyses. 

The gain in tracking efficiency obtained using per-TDC-chip $t_{0}$ constants rather
than per-module constants has been estimated with 
simulated $B^{+}\to J/\psi K^{+}$ decays reconstructed in the LHCb
detector. The tracking efficiency improved marginally by about 0.5 per mille per track.
The improvement in drift-time resolution and the corresponding possible new use of the time information
are discussed below.


\section{Updated drift-time distance relation and improved resolution}
\label{sec:rt-relation}

\subsection{TR-relation}

\begin{figure}[!ht]
    \centering
	\includegraphics[width=0.55\textwidth]{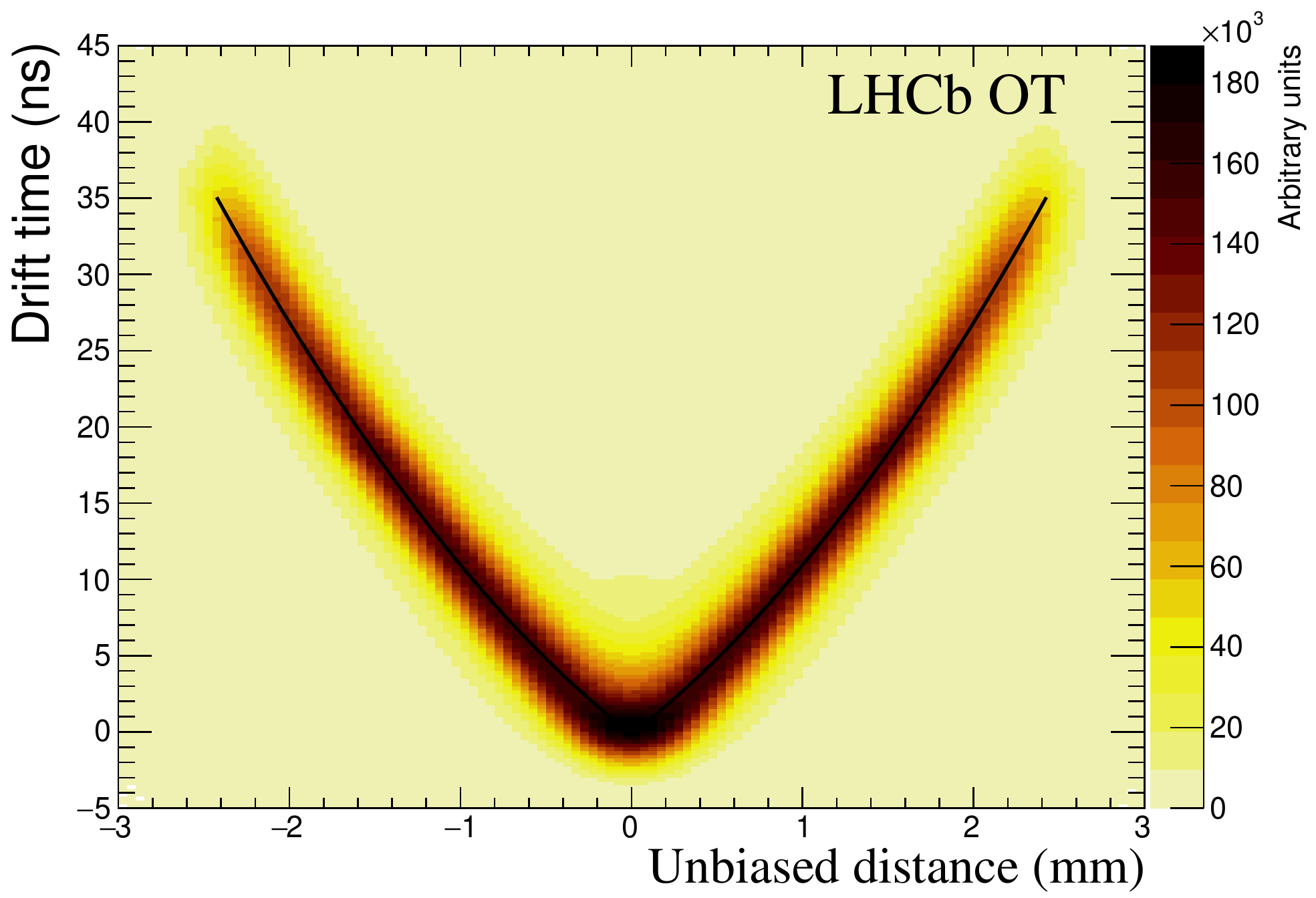}
    \caption{The drift time versus the unbiased distance distribution with the overlaid TR-relation curve, 
      obtained from the fit (black line).}
    \label{fig:tr_relation}
\end{figure}

The drift-time measured by the detector needs to be converted to a distance from the wire to be used 
in the tracking. 
Drift-time and position information are related by means of the drift-time-distance relation, or TR-relation.
The TR-relation is calibrated on data using good quality tracks, selected by requiring a momentum larger than $10 \gevc$ 
and a track-fit $\chi^{2}/\text{ndf} < 2$; the track-fit is unbiased by excluding the hit under study.  
The parametrization of the TR-relation is extracted from the fit to the distribution of the measured drift-time 
versus the distance of the reconstructed track from the wire, and is shown in figure~\ref{fig:tr_relation}.
The TR-relation is parametrised with a second order polynomial as a function of the unbiased distance.
The TR-relation with parameter values obtained with Run 2 data is 
\begin{equation} 
	t(r) = \left(21.3 \frac{\vert r \vert}{R} + 14.4 \frac{\vert r \vert^{2}}{R^{2}} \right) \ns ,
\end{equation}
where $r$ is the distance and $R$ is the radius of the straw-tube. The parameter values  
are consistent with the values observed in Run 1~\cite{Arink:1629476}.

\subsection{Resolution}

The time resolution dependence on the distance from the wire is extracted from the 2-dimensional 
fit to the distribution in figure~\ref{fig:tr_relation} to be
\begin{equation}
\label{eq:res}
	\sigma_{t}(r) = \left(2.25 + 0.3 \frac{\vert r \vert}{R} \right) \ns \quad.
\end{equation}
The average drift-time resolution is obtained from equation~\ref{eq:res} to be $2.40 \ns$, 
which is an improvement of about $20 \%$ with respect to during Run 1.
 Similarly, the spatial resolution
is reduced from $205 \mum$ to $171 \mum$ due to the combination of monolayer
alignment, global drift time calibration and single FE-module calibration,
discussed in the previous sections.
\begin{figure}[!ht]
    \centering
    \includegraphics[width=0.49\textwidth]{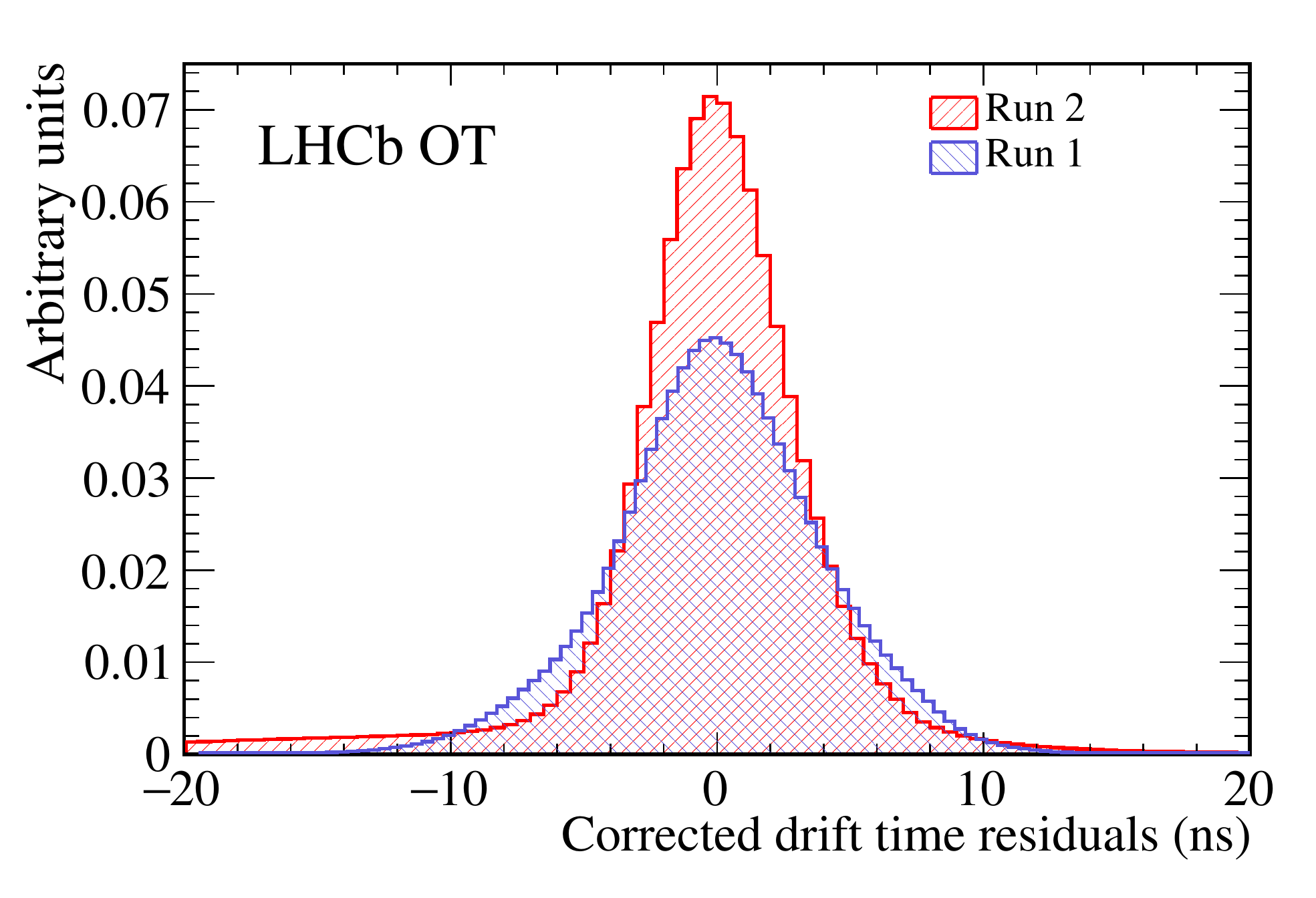}\put(-175,100){(a)}
    \includegraphics[width=0.49\textwidth]{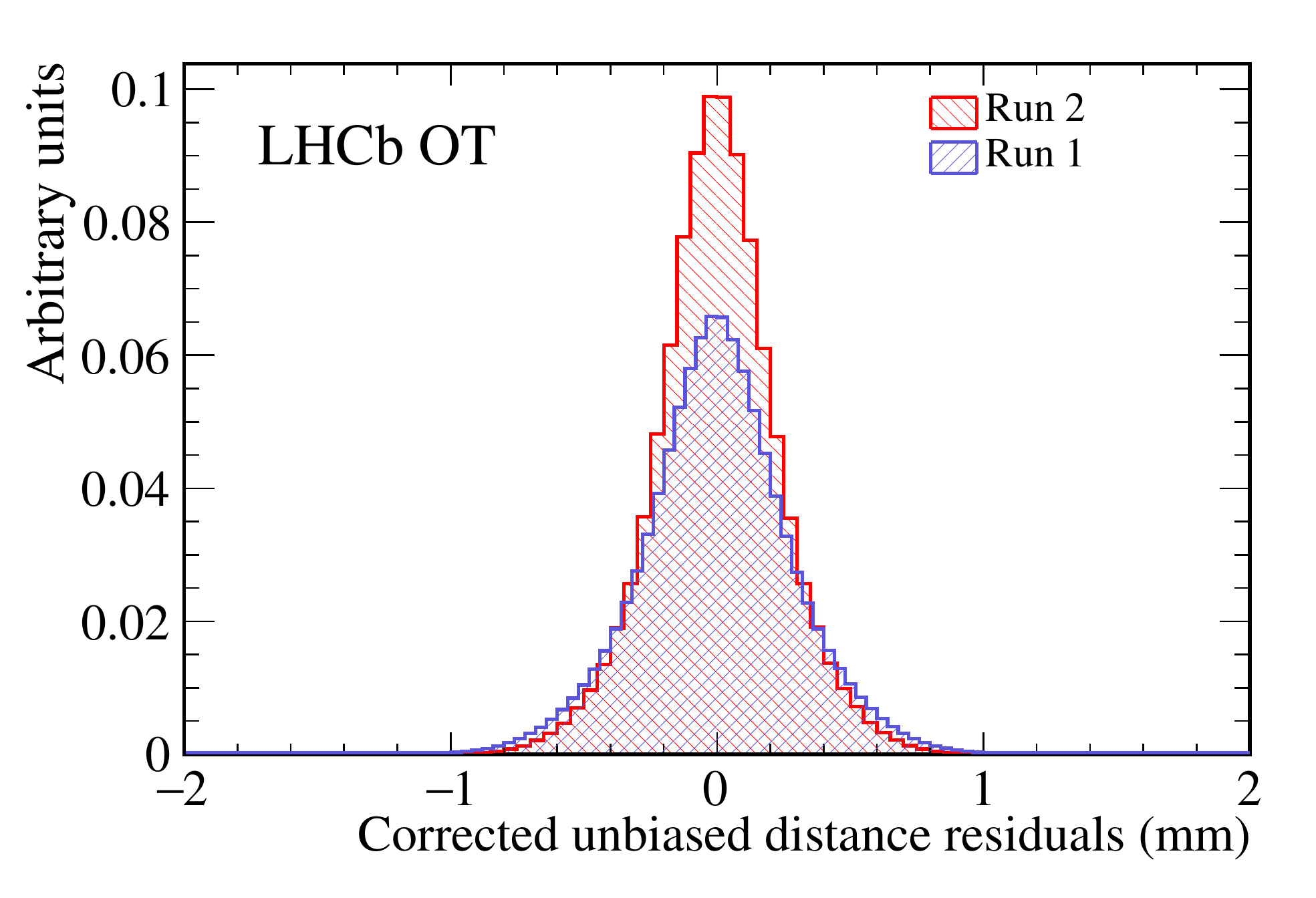}\put(-175,100){(b)}
    \caption{ (a) Drift-time residual distribution and (b) hit distance residual distribution.}
    \label{fig:resolution}
\end{figure}
The time and spatial distribution comparing the measured
drift-time and hit position of Run 1 and Run 2 are shown in figure~\ref{fig:resolution}  for illustration only, 
corrected for the finite precision of the track parameters.


\clearpage 
\section{Use of time information}
\label{sec:tracktimes}

The more precise time and space alignment, has led to an improved drift-time hit resolution of the OT, 
beyond its design specifications. 
This opens the possibility that timing information is assigned to physics objects, 
like reconstructed tracks and vertices.
A sample of $100\,000$ selected events containing a semileptonic $B$ meson decay
has been used for this study.

\subsection{Time-stamp for single tracks}

\begin{figure}[!b]
  \begin{center}
    \includegraphics[width=0.49\linewidth]{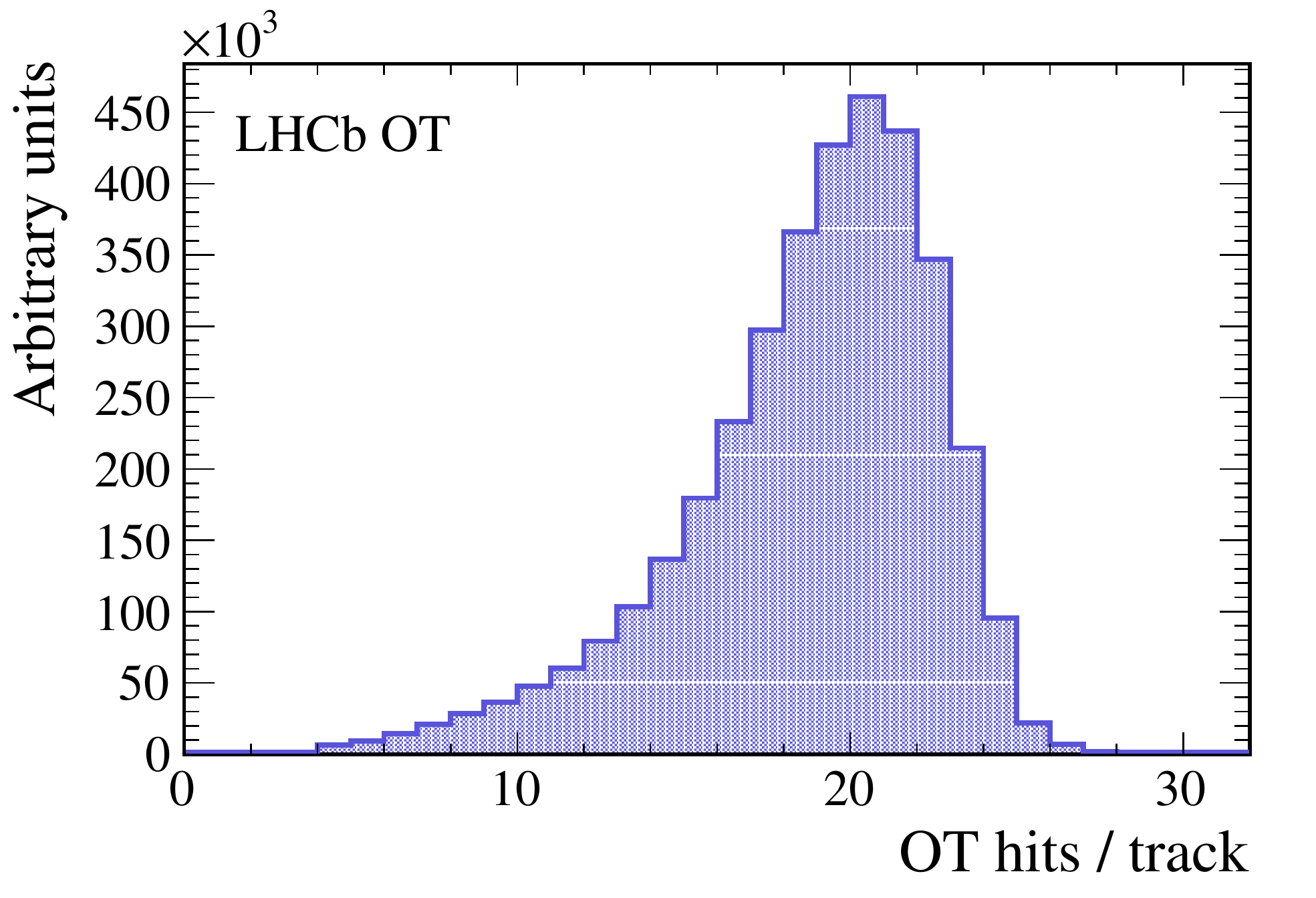}\put(-175,100){(a)}
    \includegraphics[width=0.49\linewidth]{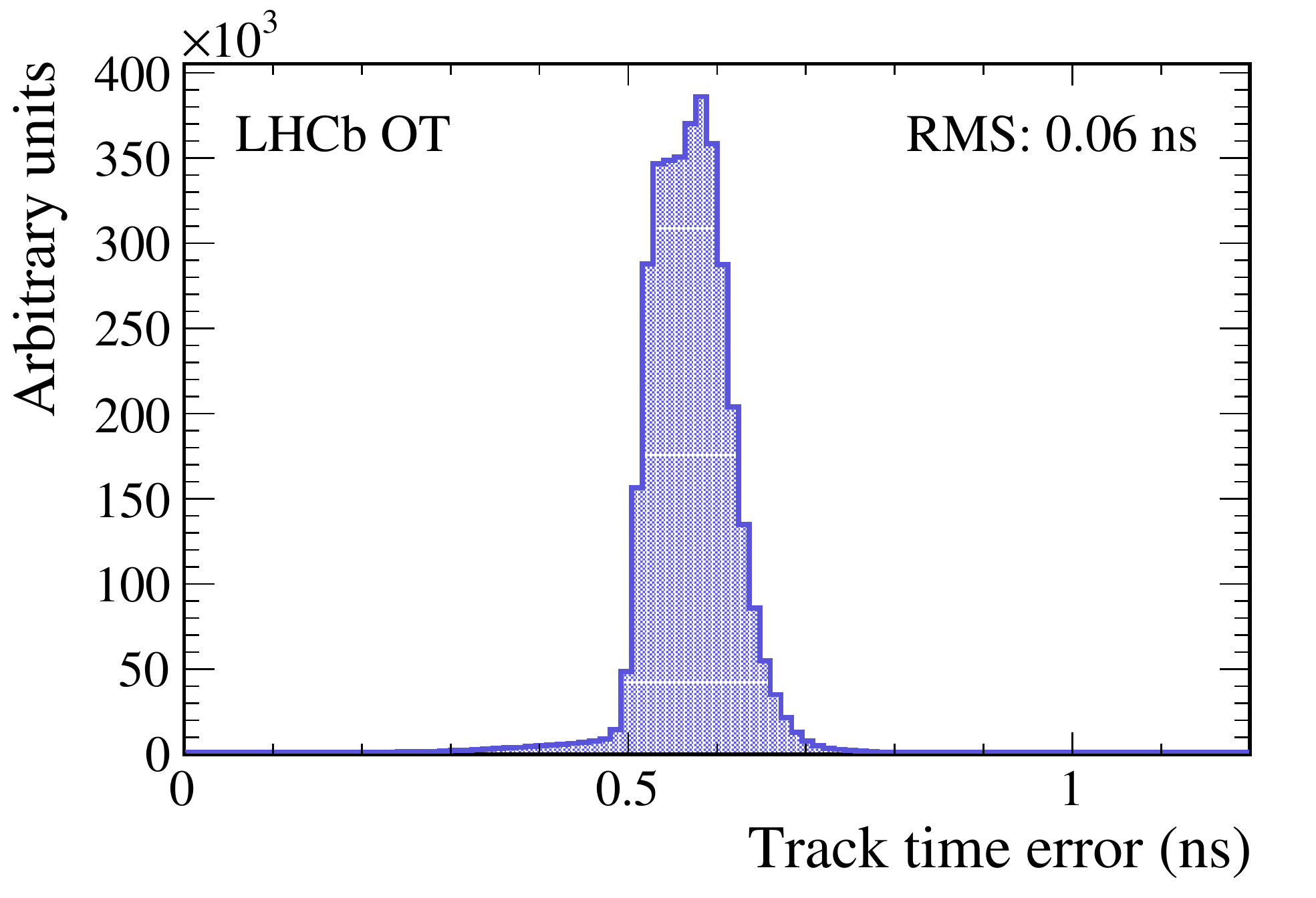}\put(-175,100){(b)}
  \end{center}
  \caption{
    \small (a) The number of OT hits per track $N$, where the drift-time is used.
   (b) The uncertainty on the time per track, as estimated from time resolution per hit, 
    $\sigma_t/\sqrt{N}$.}
  \label{fig:tracktime-delta}
\end{figure}

A single track that traverses the entire OT typically contains 22 OT hits. For each hit the
drift time residuals can be calculated, as described in
section~\ref{sec:realtime}. The track time is defined as the sum of all individual
drift time residuals, weighted by their errors. This track time is the
difference in the arrival time of the particle at the OT with respect to 
the expected arrival time for a particle travelling at the speed of light. The number of
OT hits per track used for the track time calculation is shown in
figure~\ref{fig:tracktime-delta}(a). The mean number of hits is about 18, with a
long tail to the left. The actual number of OT hits on a track is slightly larger, due to the addition
of OT hits where the drift time is not used (for example for straws that are hit twice, and for which
the drift time does not agree with the track position).
Tracks with a number of hits less than three are not considered in these studies,
and neither are tracks with a momentum below $2~\gevc$. 
The uncertainty of the track time is obtained from the drift time residual, $\sigma_t$,
and is thus expected to scale as  $\sigma_t/\sqrt{N}$, where $N$ is the number of hits for the given track.
The distribution is shown in
figure~\ref{fig:tracktime-delta}(b), and has a mean value of $0.57~\ns$. 

A track reconstructed with random OT hits (e.g. from noise or spill-over) is called a ghost track. The
track time of ghost tracks is expected to be large, and could thus be of added
value in their rejection. Studies on data show that the track time
of ghost tracks has indeed discriminating power, 
however it was also found that it is largely correlated with other
ghost-reducing parameters, such as the $\chi^2$ of the track fit, and thus adds little value.

\subsection{Time-of-flight for pions and protons}

\begin{figure}[!b]
  \centering
    \includegraphics[width=0.7\linewidth]{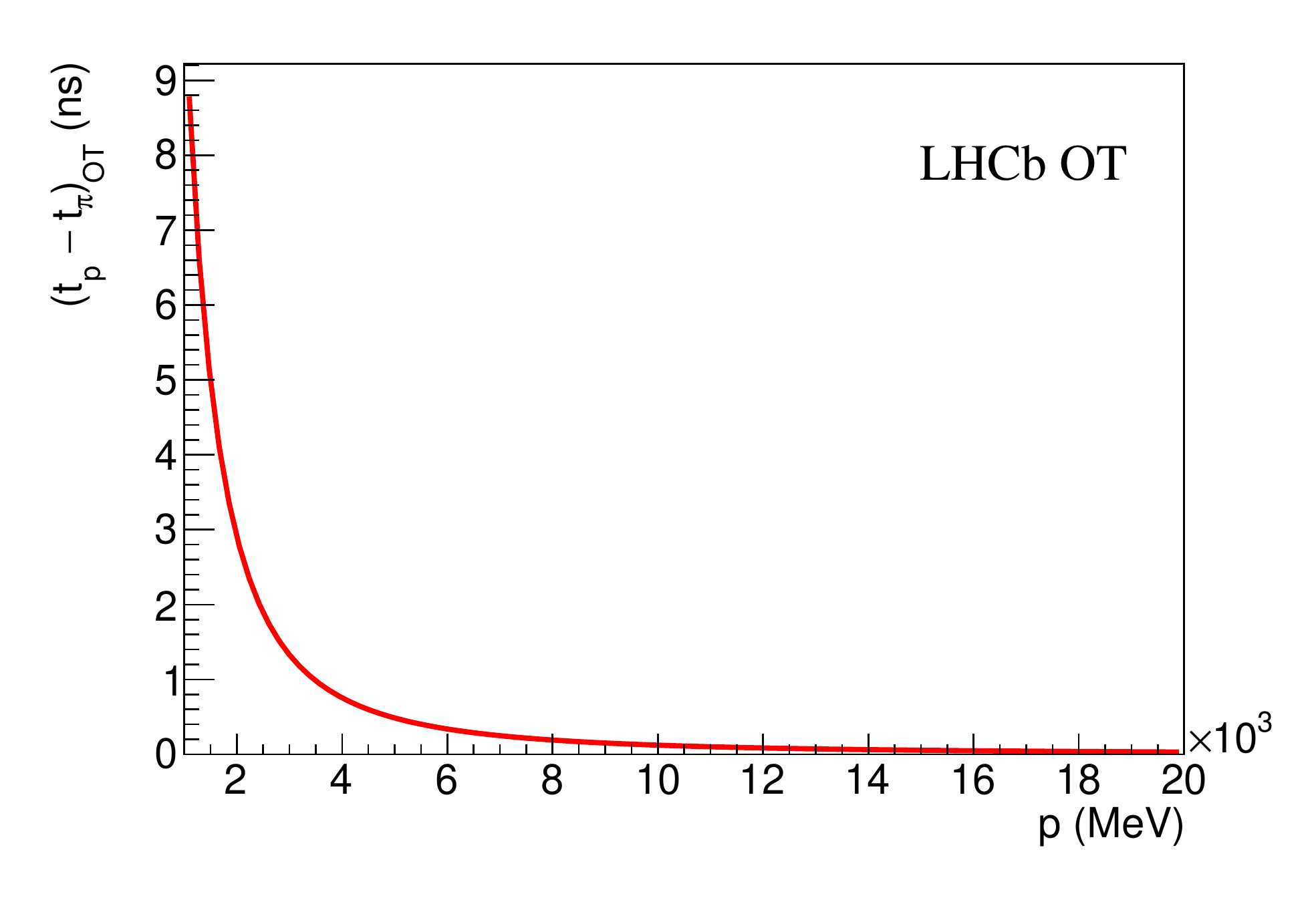}
  \caption{
    \small The difference in time-of-flight between protons and pions as a function of their momentum.}
  \label{fig:tracktime-tof-diff}
\end{figure}

The velocity of particles created in $pp$ collision can mostly be
approximated by the speed of light. However, heavy, low momentum particles have
a lower velocity, which can lead to  a significant later
arrival time. For protons, this is about $0.5\ns$ at $5\gevc$ at the
centre of the OT - about $8.5 \m$ from the interaction point - as shown in
figure~\ref{fig:tracktime-tof-diff}. This is similar to the expected track time
resolution. 

\begin{figure}[!ht]
  \begin{center}
    \includegraphics[width=0.49\linewidth]{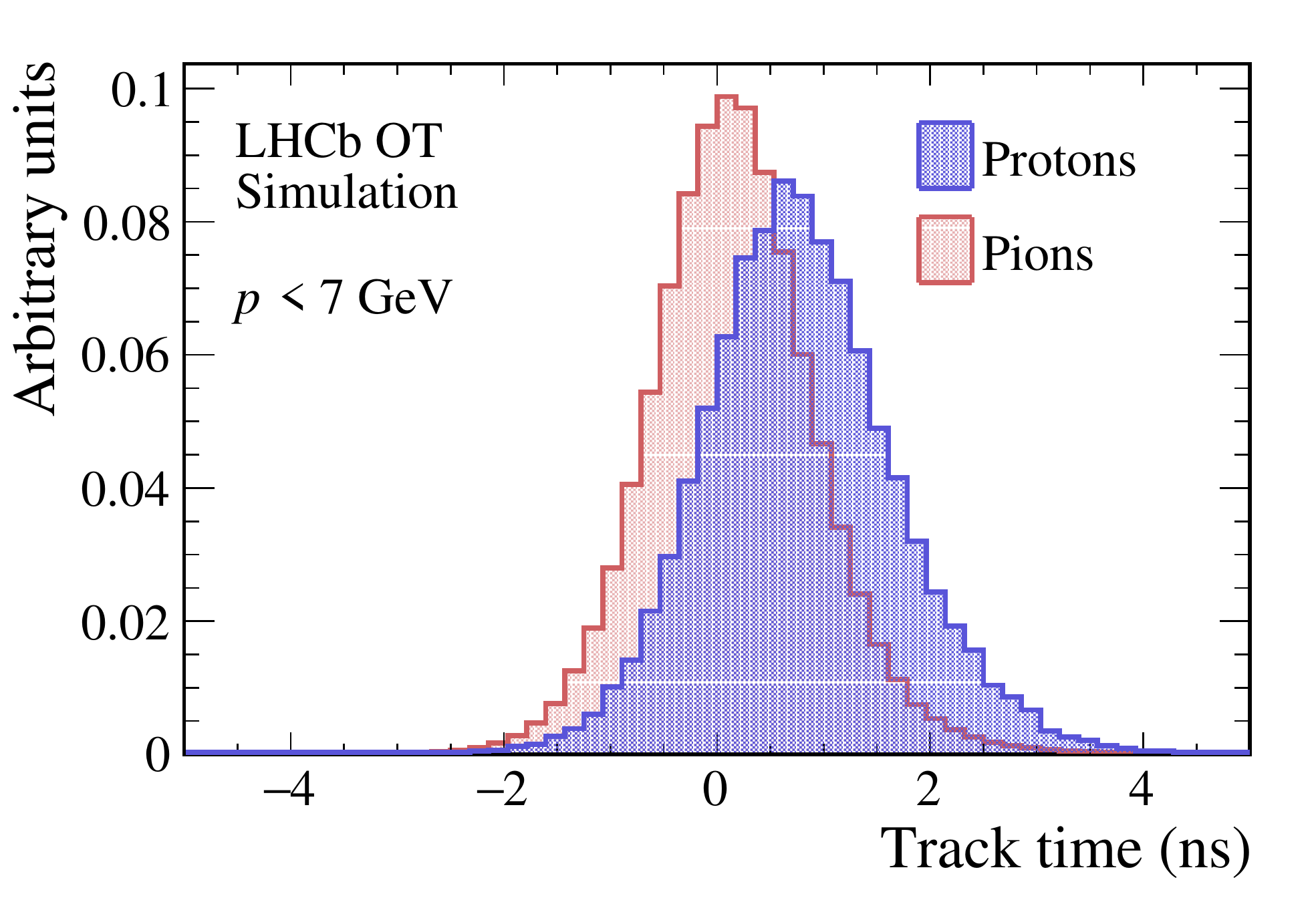}\put(-177,80){(a)}
    \includegraphics[width=0.49\linewidth]{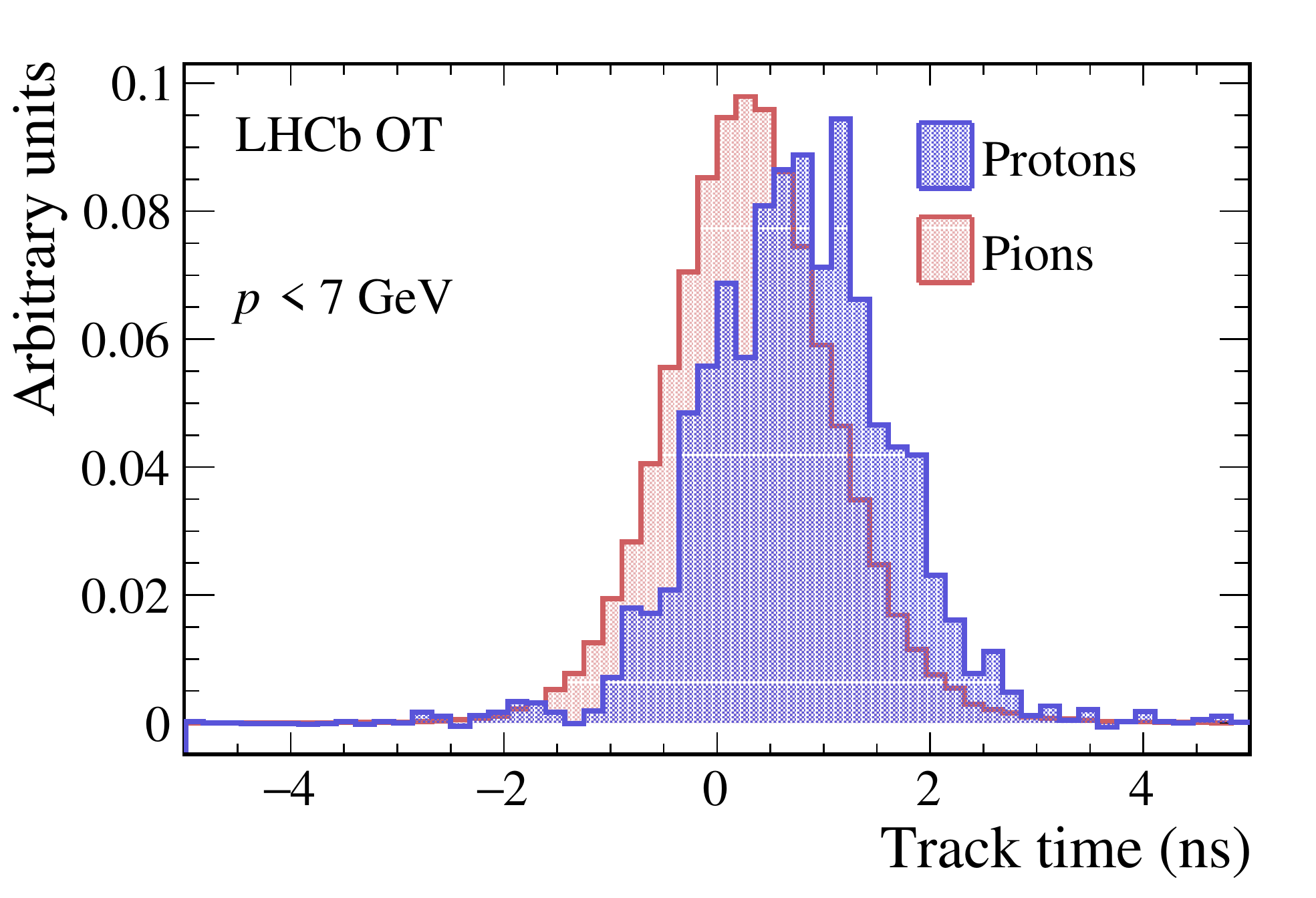}\put(-177,80){(b)}
  \end{center}
  \caption{
    \small The distribution of track times for protons and pions with $p<7 \gevc$ in (a) simulation and (b) in data.}
  \label{fig:tracktime-tof}
\end{figure}

\begin{figure}[!hb]
  \begin{center}
    \includegraphics[width=0.49\linewidth]{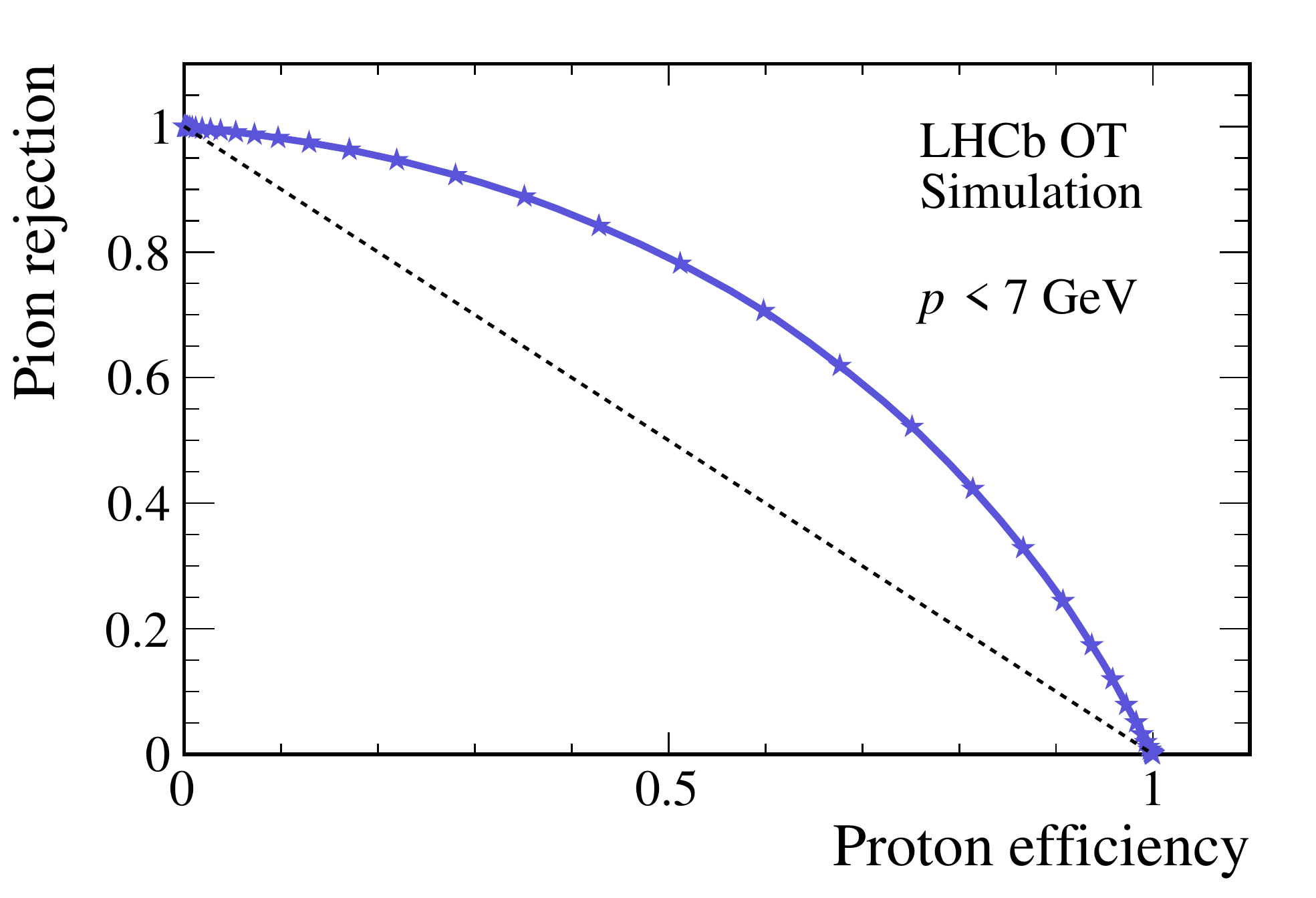}\put(-175,100){(a)}
    \includegraphics[width=0.49\linewidth]{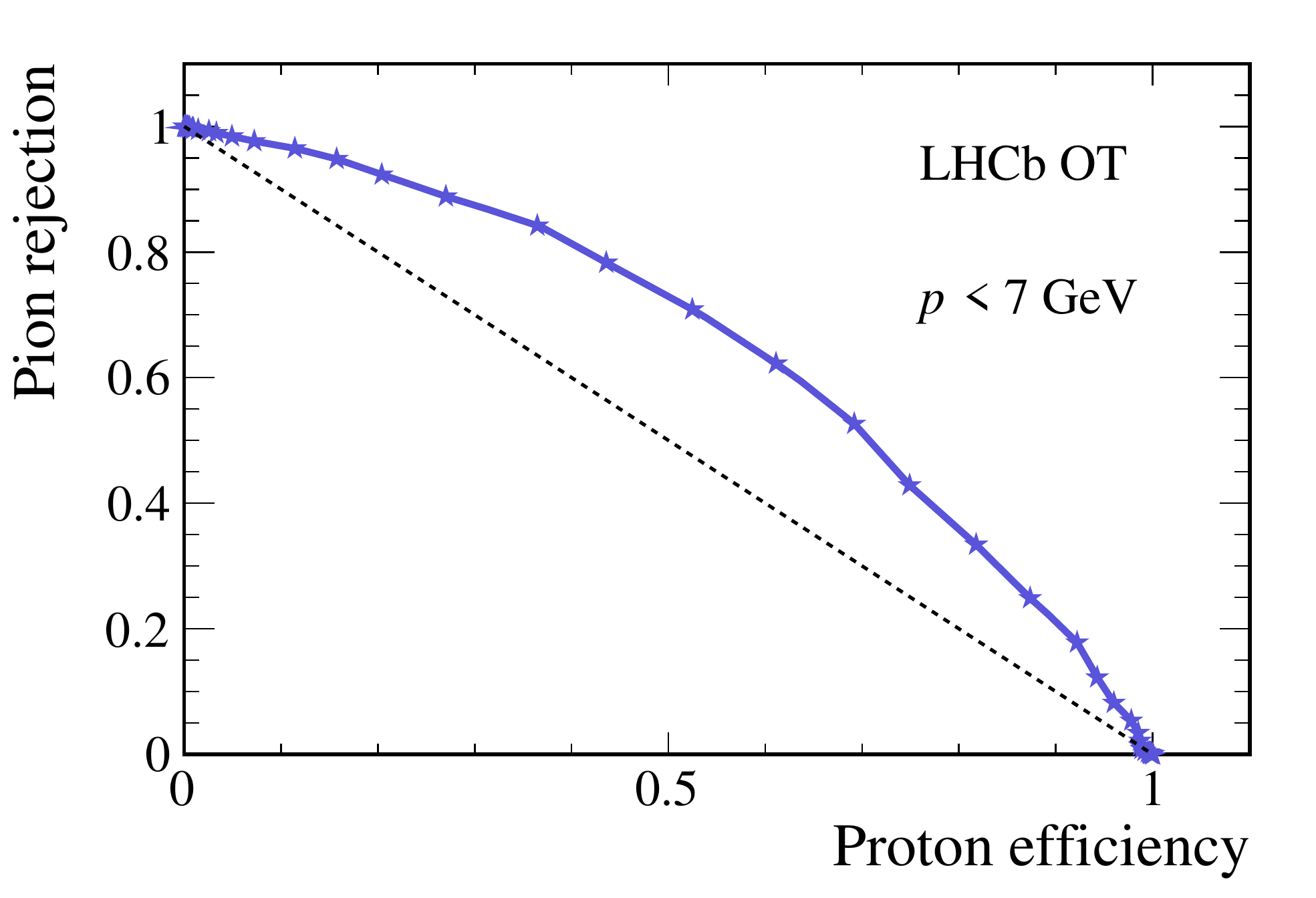}\put(-175,100){(b)}
  \end{center}
  \caption{
    \small The pion rejection rate as a function of the proton efficiency for (a) simulation and (b) data.}
  \label{fig:tracktime-roc}
\end{figure}

The distribution of track times of pions and protons with momenta below $7~\gevc$ in \lhcb
simulation is shown in figure~\ref{fig:tracktime-tof}(a). In data, a sample of
\Dstar tagged $\Dz \to \Kp \pim$ decays is used as a source of unbiased
identified pions. This sample is regularly used to calibrate the \lhcb particle
identification response, since pions can be identified using only their charge.
In a similar fashion, events with a semileptonic \Lb decay are used as a control sample
of protons, albeit that the initial selection discards a large fraction of
low-momentum protons. The track time distribution in data is shown for both pions and protons in
figure~\ref{fig:tracktime-tof}(b). 
In both simulation and data the difference in the track time distributions
between low momentum pions and protons is clearly visible. The discriminating power of
the track time to distinguish protons from pions is shown in figure~\ref{fig:tracktime-roc}.
The performance in data and simulation are found to be in good agreement.
As an example, for a proton identification efficiency around 75\%, 
half of the pions are rejected.

Most of the particle identification performance of \lhcb is due to the ring
imaging Cherenkov detectors (RICH), with radiators \cffour and \cfourften. The
particle identification for momenta between $1.5$ and $10 \gevc$ was done with the use of the aerogel radiator,
but it has been removed after Run~1 fitting the design for higher luminosities.
The OT track time can assist in this momentum range in Run~2, especially for
analyses involving protons, $\Sigma$ hyperons, deuterons, and new long-lived
heavy particle searches. 
In \lhcb, particle identification is done by creating a neural network using
inputs from the RICH, calorimeters and muon detector systems~\cite{Anderlini:2202412}. 
In the near future the OT track time will be added as an input variable.
The addition of the OT track-time will improve also the performance of particle identification 
of low-momentum particles used to identify the flavour of $B$-mesons (flavour tagging).
Finally, the particle identification performance of the OT track time serves as
a test case for a possible future time-of-flight detectors
for LHC, such as for example TORCH~\cite{Charles:2010at}.

\subsection{Time-stamp of primary vertices}

\begin{figure}[!h]
  \begin{center}
    \includegraphics[width=0.49\linewidth]{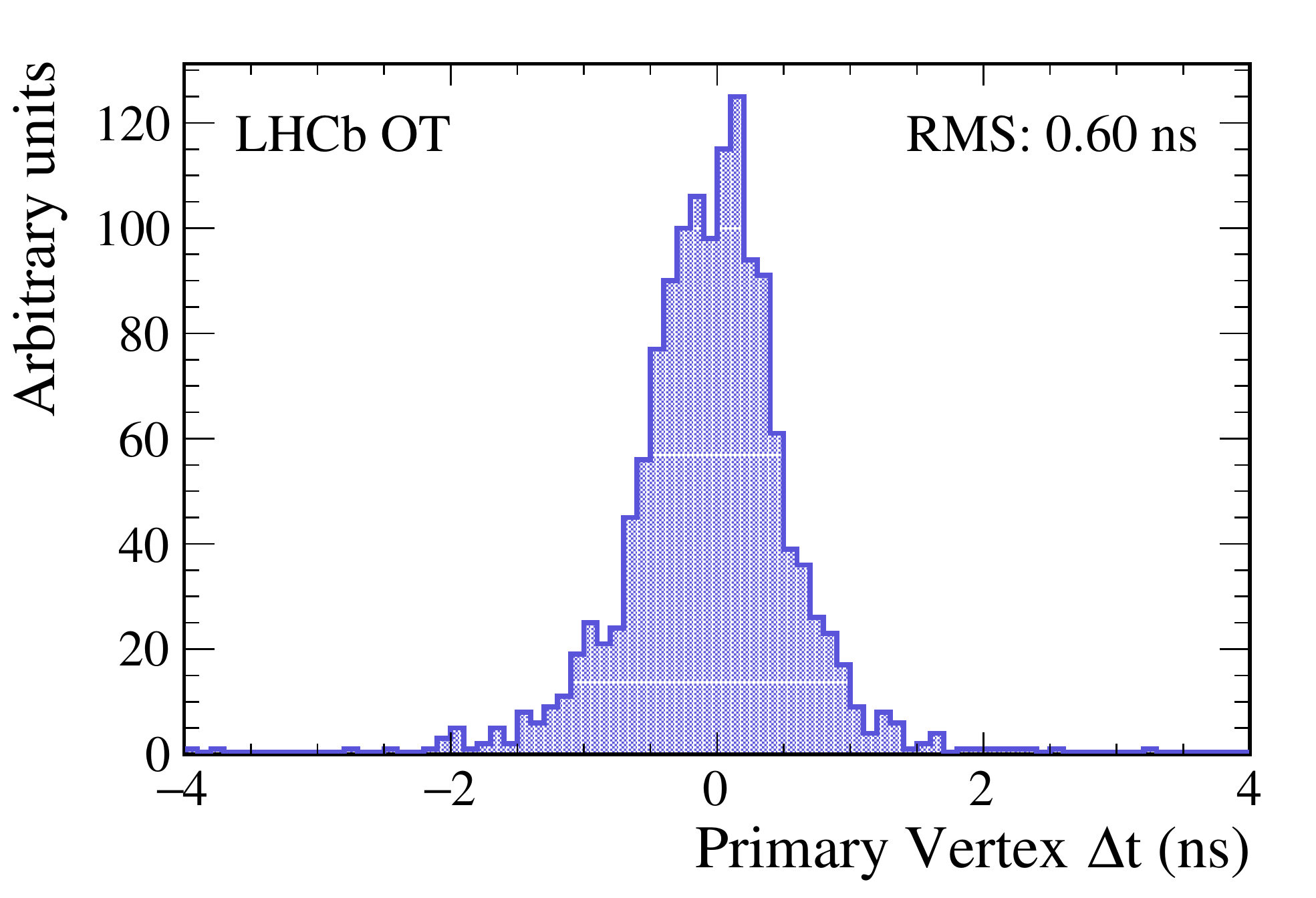}\put(-175,100){(a)}
\includegraphics[width=0.49\linewidth]{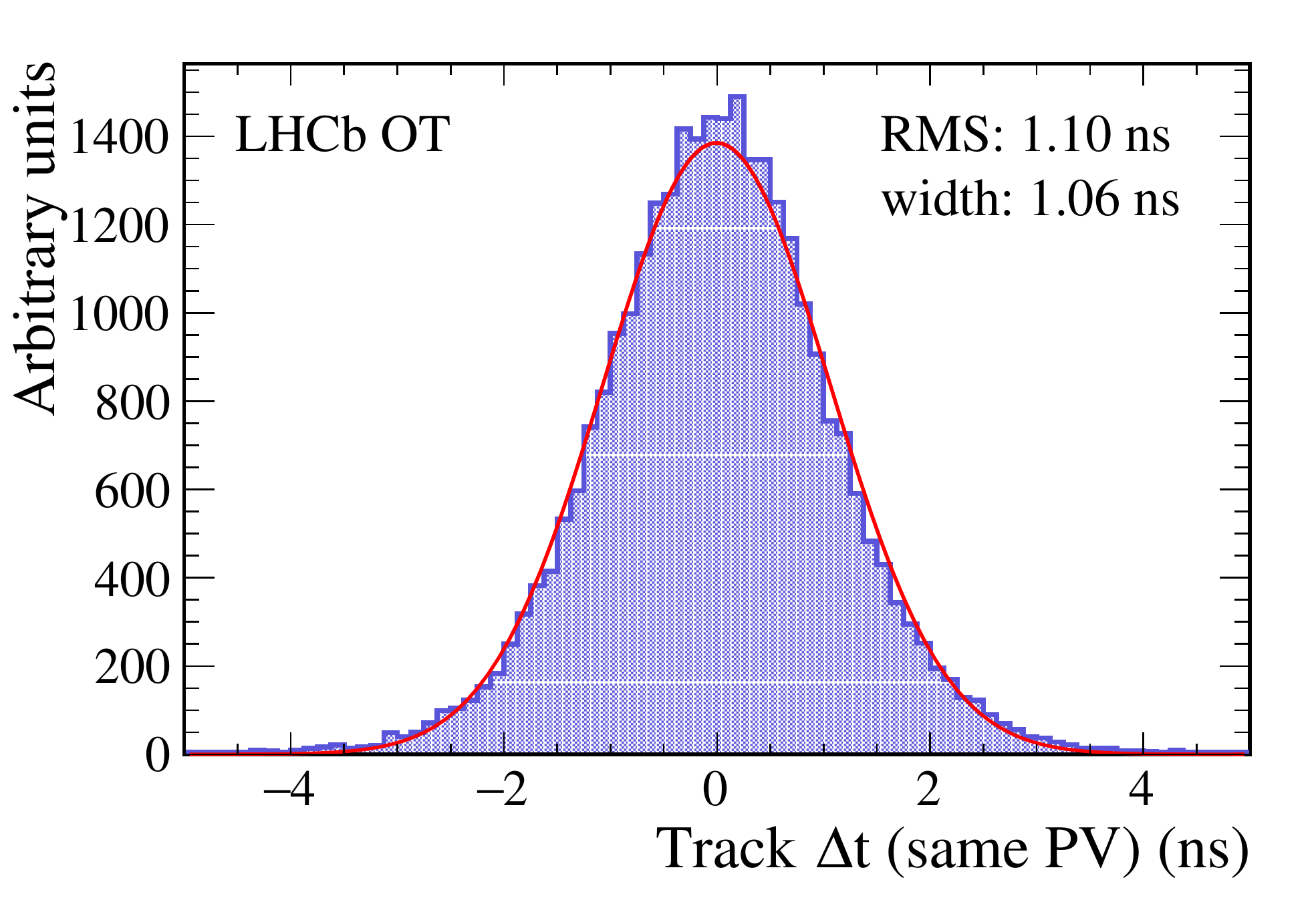}\put(-175,100){(b)}
  \end{center}
  \caption{
    \small (a) The difference between the average time per primary vertex, for events with two reconstructed primary vertices.
   (b) The difference between the time of two tracks belonging to the same primary vertex. 
   }
  \label{fig:tracktime-PV}
\end{figure}

One of the experimental limitations to increase the luminosity at \lhcb is the number of $pp$ collisions
per bunch crossing, which hampers the association of a $B$-decay to the correct primary vertex.
Two primary vertices (PVs) separated far enough in space can be resolved by the
excellent spatial resolution of the vertex reconstruction. A time stamp per PV would aid in the
separation of these vertices.
The PV time is calculated as the weighted average of the track times of all
tracks associated with that PV. 
For events with two PVs, the PV time difference is shown in
figure~\ref{fig:tracktime-PV}(a). 
The expected error for this spread is $\sigma_{t,track} / \sqrt{N_{tracks}}$, 
where $N_{tracks}$ is the number of tracks per PV, which is 25 on average. 
The significance of this spread when comparing
to the expected error is found to be larger than one, indicating some
discriminating power. However, it is also larger than the combined expected PV time uncertainty ($0.57\ns/\sqrt{25}$) 
with the PV spread of $0.18 \ns$ that is expected from the \lhc~\cite{privArduini}.

In figure~\ref{fig:tracktime-PV}(b) the difference in track time between two
tracks from the same PV is shown. The RMS is larger than the expected value of
$0.57 \times \sqrt{2} \ns$. This indicates that the track time errors are
underestimated by about 20\%.

\section{Conclusions}
\label{sec:conclusions}

In this article the performance of the LHCb Outer Tracker detector
has been summarized as obtained from LHC Run 2 data. 

The OT has reached and in some cases superseded the design performances, 
being able to acquire data in $pp$ collisions at luminosities twice higher 
than anticipated and to be run in the high multiplicity environment of heavy ion collisions. 

Efficiencies and availability of the detector components have been kept at maximal level
for the entire data taking period, 
and no signs of ageing or damages due to radiation have been observed. 

Novel methods have been implemented for the time and space alignment, 
which have improved the resolution of the detector.
This allowed the study to apply the time information beyond the tracking purposes. 
A time-of-flight measurement has been developed providing 
discrimination for particle identification of low momentum charged hadrons, 
and that will be helpful in physics analyses with Run 2 data. 
This offers also benchmarking possibilities for future time-sensitive detectors for 
particle identification or multiple-interactions event tagging.


\section*{Acknowledgements}
 
\noindent We express our gratitude to our colleagues in the CERN
accelerator departments for the excellent performance of the LHC. We
thank the technical and administrative staff at the LHCb
institutes. We acknowledge support from CERN and from the national
agencies: CAPES, CNPq, FAPERJ and FINEP (Brazil); NSFC (China);
CNRS/IN2P3 (France); BMBF, DFG and MPG (Germany); INFN (Italy); 
FOM and NWO (The Netherlands); MNiSW and NCN (Poland); MEN/IFA (Romania); 
MinES and FANO (Russia); MinECo (Spain); SNSF and SER (Switzerland); 
NASU (Ukraine); STFC (United Kingdom); NSF (USA).
We acknowledge the computing resources that are provided by CERN, IN2P3 (France), KIT and DESY (Germany), INFN (Italy), SURF (The Netherlands), PIC (Spain), GridPP (United Kingdom), RRCKI (Russia), CSCS (Switzerland), IFIN-HH (Romania), CBPF (Brazil), PL-GRID (Poland) and OSC (USA). We are indebted to the communities behind the multiple open 
source software packages on which we depend. We are also thankful for the 
computing resources and the access to software R\&D tools provided by Yandex LLC (Russia).
Individual groups or members have received support from AvH Foundation (Germany),
EPLANET, Marie Sk\l{}odowska-Curie Actions and ERC (European Union), 
Conseil G\'{e}n\'{e}ral de Haute-Savoie, Labex ENIGMASS and OCEVU, 
R\'{e}gion Auvergne (France), RFBR (Russia), XuntaGal and GENCAT (Spain), The Royal Society 
and Royal Commission for the Exhibition of 1851 (United Kingdom).


\addcontentsline{toc}{section}{References}
\setboolean{inbibliography}{true}
\bibliographystyle{LHCb}
\bibliography{main,LHCb-PAPER,LHCb-CONF,LHCb-DP,LHCb-TDR}

\end{document}